\documentclass[aps,pra,reprint,groupedaddress]{revtex4-2}

\usepackage{graphicx}
\usepackage{color}
\usepackage{mathtools}

\usepackage{tikz}
\usepackage[percent]{overpic}

\begin{document}



\title{Correspondence between the surface integral and linear combination of atomic orbitals methods for ionic-covalent interactions in mutual neutralisation processes involving H$^-$/D$^-$}


\author{Paul S. Barklem}
\affiliation{Theoretical Astrophysics, Department of Physics and Astronomy, Uppsala University, Box 516, SE-751 20 Uppsala, Sweden}


\date{\today}

\begin{abstract}
The surface integral method for estimating ionic-covalent interactions in diatomic systems been successful in producing cross sections for mutual neutralisation (MN) in reasonable agreement with experimental results for branching fractions between final states in systems such as O$^+$/O$^-$ and N$^+$/O$^-$.  However, for simpler cases of MN involving H$^-$ or D$^-$, such as Li$^+$/D$^-$ and Na$^+$/D$^-$, it has not produced results that are in agreement with experiments and other theoretical calculations; in particular, for Li$^+$/D$^-$ calculations predict the wrong ordering of importance of final channels, including the incorrect most populated channel.  The reason for this anomaly is investigated and a leading constant to the asymptotic H$^-$ wavefunction is found that is different by roughly a factor $1/\sqrt{2}$ to that which has been used in previous calculations with the surface integral method involving H$^-$ or D$^-$.  With this correction, far better agreement with both experimental results and with calculations with full quantum and LCAO methods is obtained. Further, it is shown that the surface integral method and LCAO methods have the same asymptotic behaviour, in contrast to previous claims.  This result suggests the surface integral method, which is comparatively easy to calculate, has greater potential for estimating MN processes than earlier comparisons had suggested.
\end{abstract}


\maketitle

\section{Introduction}

Charge transfer, where an electron moves from one atom or ion to another during a collision, is a fundamental atomic process. Mutual neutralisation (MN), with corresponding reverse process ion-pair production
\begin{equation}
\mathrm{A}^+ + \mathrm{B}^- \rightleftharpoons \mathrm{A}^* + \mathrm{B}
\label{eq:mn}
\end{equation}
is an important special case, and methods for estimating cross sections are important for interpreting experimental results and testing our understanding of the physical mechanisms involved \citep{LaunoyMutualNeutralizationLi2019, EklundCryogenicmergedionbeamexperiments2020, EklundFinalstateresolvedmutualneutralization2021}, as well as modelling various plasma environments such as planetary and planetary satellite atmospheres \cite{VuittonNegativeionchemistry2009} including the earth's ionosphere \cite{SagawaLongitudinalstructureequatorial2005}, stellar atmospheres \cite{Barklem2003b, BarklemMutualNeutralizationLi2021}, as well as being important for diagnostics in and fusion plasma applications and high energy physics experiments employing negative ion sources \cite{Fantznoveldiagnostictechnique2006}. The transferred electron is captured into specific excited states $\mathrm{A}^*$, and thus affects the distribution of populations of states of atom A and its observed spectrum.  These processes occur predominantly through interactions between ionic ($\mathrm{A}^+ + \mathrm{B}^-$) and covalent ($\mathrm{A}^* + \mathrm{B}$) configurations at avoided crossings in the adiabatic potential energy curves e.g.~\cite{zener_nonadiabatic_1932, Bates14TheoreticalTreatment1962}.  Any method for estimating MN cross sections thus hinges on the ability to calculate the ionic-covalent interactions. 

A promising approach is the surface integral method.  The general technique was independently developed roughly simultaneously by a number of workers, and thus is known by various names including the Holstein-Herring method, e.g.\ \cite{ScottExchangeenergyH21991}, the Firsov-Landau-Herring method, e.g.\ \cite{Chibisov1988}, and the Landau-Herring method, e.g.\ \cite{DavidovicResonantChargeExchange1969,Janev1972, Janev1976}.  The original papers on the idea include those by Firsov \cite{Firsov1951}, Holstein \cite{HolsteinMobilitiesPositiveIons1952}, and Herring \cite{Herring1962}, and the method was used to solve problems in Landau and Lifshitz's text \emph{Quantum Mechanics} \cite{landau_quantum_1965}, published in Russian in the 1950's; the asymptotic (large internuclear distance $R$) splitting between gerade and ungerade states for H$_2^+$ is solved in \S81 of Ref.~\cite{landau_quantum_1965}, building on problem 3 in \S50.  The basic idea is that the interaction energy can be calculated from the electronic current flowing across a surface between the two nuclei, obtained by manipulating Schr\"odinger equations corresponding to cases where the electron is localised on either nuclei. On the assumption that the wavefunctions are close to zero at the surface, i.e. in the limit of large internuclear separations, the method allows the interaction to be derived in terms of the atomic electronic wavefunctions at the surface.  

A key point in the use of the method is the choice of the integration surface to ensure that wavefunctions are small.  Smirnov \cite{smirnov_formation_1965} derived expressions for the interaction in the $\mathrm{A}^+ + e + \mathrm{B}$ system, assuming the interaction between $e$ and B to be zero, and taking the integration surface as the midplane, halfway between the two nuclei.  Janev \& Salin \cite{Janev1972, Janev1976} instead account for the potential due to B, and choose the integration surface as a sphere around B where the potential becomes suitably small; see also Ref.~\cite{DavidovicResonantChargeExchange1969} for an earlier application in resonant charge transfer.

Janev \& Salin's expressions for the couplings from the surface integral method, which for consistency with previous work \cite{barklem_excitation_2016, barklem_erratum:_2017} we will call the Landau-Herring-Janev (LHJ) method, have been used in various studies of MN involving H$^-$, coupled with Landau-Zener dynamics, including MN of alkalis with H$^-$ \cite{Janev1978}, H$_3^+$ with H$^-$ \cite{Janev2006}, Si$^+$ with H$^-$ \cite{Jian-GuoMutualrecombinationslow2006}, H$^+$ and Be$^+$ with H$^-$ \cite{MiyanoHedberg2014}.  In Refs.~\cite{barklem_excitation_2016, barklem_erratum:_2017} the method was compared with full quantum and linear combination of atomic orbitals (LCAO) calculation results for MN of Li$^+$, Na$^+$, and Mg$^+$ with H$^-$, and the results found to differ significantly.   Recent experimental results for branching fractions in Li$^+$/D$^-$ \cite{LaunoyMutualNeutralizationLi2019, EklundCryogenicmergedionbeamexperiments2020} and Na$^+$/D$^-$ \cite{EklundFinalstateresolvedmutualneutralization2021} have allowed calculations to be tested, and full quantum and LCAO results \cite{BarklemMutualNeutralizationLi2021} have shown good agreement, implying the LHJ method is in disagreement with these experiments.  This is puzzling, given that the LHJ method has been shown to produce estimates in reasonable agreement with experiment for MN of N$^+$/O$^-$ and O$^+$/O$^-$ \cite{zhou_mutual_1997, deRuetteMutualNeutralizationSubthermal2018}.

Asymptotic methods such as LHJ and LCAO are of significant importance for many applications, as estimates can be made rather inexpensively compared to full quantum calculation (quantum chemistry potentials and couplings with quantum scattering), and thus it is valuable to resolve the origin of this discrepancy.  The fact that the LHJ method appears to work well for MN involving O$^-$, and less well for the more simple H$^-$, would seem to suggest the following possible explanations: 1) there is a problem with the H$^-$ wavefunction used, 2) the agreement for O$^-$ is fortuitous, or 3) the full quantum, LCAO and experimental results are in error.  In this paper it will be shown that an error appears to have been made in deriving the asymptotic H$^-$ wavefunction used in all applications to MN involving H$^-$, and correcting this error brings the LHJ method into reasonable agreement with other theory and experiments.  Previous claims that the discrepancy between the LHJ and LCAO methods is due to fundamental problems in the LCAO method are reexamined in light of this.





\section{Theory}

\begin{figure}
\center
\begin{tikzpicture}[scale=5]
\path (0,0) coordinate (A);
\path (1,0) coordinate (B);
\path (0.4,0.3) coordinate (1);

\path (A) ++(0,-0.1) coordinate (Alow);
\path (B) ++(0,-0.1) coordinate (Blow);

\draw [fill](A) circle (0.03);
\draw [fill](B) circle (0.03);
\draw (A) circle (0.10);
\draw (B) circle (0.10);
\draw [fill](1) circle (0.005);
\draw (A) -- (B);
\draw (A) -- (1);
\draw (B) -- (1);

\node [below] at (Alow) {A$^+$};
\node [below] at (Blow) {B};
\node [above] at (1) {$e$};

\node [left] at (0.35, 0.15) {$r_{\mathrm{A}}$};
\node [left] at (0.65, 0.15) {$r_{\mathrm{B}}$};
\node [above] at (0.5, 0.0) {$R$};

\draw [dashed] (A) circle (0.2);
\draw [dashed] (B) circle (0.2);
\node [below] at (-0.2, 0.2) {$c$};
\node [below] at (1.2, 0.2) {$i$};

\end{tikzpicture}
\caption{Sketch of the $\mathrm{A}^+ + e + \mathrm{B}$ system, where $\mathrm{A}^+$ is the singly charged core, $e$ the active electron, and B a neutral atom.  In the MN process (eqn.~\ref{eq:mn}), the electron moves from being localised on centre B, the ionic configuration $i$, to being localised on A, the covalent configuration $c$.}
\label{fig:system}
\end{figure}
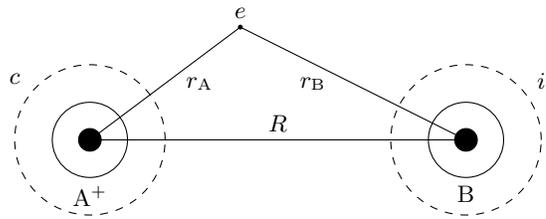

Mutual neutralisation occurs predominantly at avoided crossings between adiabatic molecular potential curves, corresponding to real crossings in a diabatic representation, e.g. \cite{zener_nonadiabatic_1932,Bates14TheoreticalTreatment1962}.  In a diabatic representation the states of the system can be described in terms of ionic and covalent configurations.  The ionic configuration is the case where the active electron is located on core B, that is $\mathrm{A}^+ + (e + \mathrm{B})$, labelled $i$ with corresponding diabatic electronic wavefunction $\Phi_{i}$, and the covalent configuration the case where the active electron is located on the core A$^+$, that is $(\mathrm{A}^+ + e) + \mathrm{B}$, labelled $c$ with wavefunction $\Phi_{c}$; see Fig.~\ref{fig:system} for a sketch of the system.  These two configurations have corresponding diabatic states with potentials that cross, $H_{ii}=\langle \Phi_{i} | H | \Phi_{i} \rangle$ and $H_{cc}=\langle \Phi_{c} | H | \Phi_{c} \rangle$ where $H$ is the Hamiltonian, and have an off-diagonal interaction (coupling) $H_{ic}=\langle \Phi_{i} | H | \Phi_{c} \rangle$.  In the adiabatic representation, the potentials avoid crossing, and a key quantity is the splitting between the adiabatic potential curves $\Delta U$, as it enters the Landau-Zener formula for the transition probability \cite{zener_nonadiabatic_1932, landau_1932, landau_1932-1}.  The transition probability between adiabatic states is (in atomic units, used throughout)
\begin{equation}
P = \exp \left(- \frac{\pi \Delta U^2}{2 v_R |d(H_{ii}-H_{cc})/dR|_{R=R_x} } \right),
\end{equation}
where $v_R$ is the radial component of velocity of relative motion of the nuclei, and $R$ is the internuclear distance and $R_x$ the distance at which the diabatic potentials $H_{ii}$ and $H_{cc}$ cross.  The splitting $\Delta U$ is related to the interaction $H_{ic}=\langle \Phi_{i} | H | \Phi_{c}\rangle$ in the (non-orthogonal) diabatic basis (e.g. \cite{AndreevExchangeinteractiontwo1973, Grice1974}) 
\begin{equation}
\Delta U(R_x) = \frac{| \Delta(R_x) |}{1-S_{ic}^2},  \; \Delta(R_x) = 2 (H_{ii}S_{ic} - H_{ic}) \big|_{R=R_x}
\label{eq:deltaU}
\end{equation}
where $S_{ic} = \langle \Phi_{i} | \Phi_{c}\rangle$ is the overlap between states $i$ and $c$. 
Note that if the diabatic basis is orthogonal 
\begin{equation}
\Delta U(R_x) = \Delta (R_x) = 2 H_{ic}(R_x).
\end{equation}
References given above, as well as for example Refs.~\cite{Bates14TheoreticalTreatment1962, Nikitin1984, barklem_excitation_2016}, can be consulted for more details on the Landau-Zener model, and the relationship between adiabatic and diabatic representations, including orthogonal and non-orthogonal representations.  

\subsection{Surface integral method}
\label{sect:surface}

The surface integral method can be used to estimate the quantity $\Delta$ via 
\begin{equation}
\Delta(R_x) = \int_s (\Phi_{i}^* \nabla \Phi_{c} - \Phi_{c}^* \nabla \Phi_{i}) d\vec{s}
\end{equation}
where $s$ is the surface of integration; see Refs.~\cite{Janev1972, Janev1976, AndreevExchangeinteractiontwo1973, Chibisov1988} for more details.  Note the wavefunctions $\Phi$ are here only the spatial components.

The LHJ theory, choosing the surface $s$ to be a sphere centred on $B$, derives an analytic expression for $\Delta(R_x)$  written in terms of the asymptotic forms of the spatial atomic wavefunction $\varphi$ of the active electron.  That is
\begin{equation}
\Phi(\vec{r}, \vec{R}) \xrightarrow[R\rightarrow \infty]{} \varphi(\vec{r}) = \mathcal{R}_{nl}(r) Y_{lm}(\theta, \phi)
\label{eq:}
\end{equation}
such that $\mathcal{R}_{nl}(r)$ is the radial part of the wavefunction, $Y_{lm}$ is the spherical harmonic function, and $\vec{r}$ is the active electron coordinate with respect to the relevant nucleus ($\vec{r_A}$ or $\vec{r_B}$).  When the active electron is on core A$^+$, i.e. the system is in the covalent configuration $c$, the asymptotic behaviour of the Coulomb wavefunction (i.e. the leading term of the asymptotic expansion) is used
\begin{equation}
\mathcal{R}_c(r) = N_c r^{1/\gamma_c - 1} \exp{(-\gamma_c r)}
\label{eq:wf_cov}
\end{equation}
where $-\gamma_c^2/2$ is the binding energy of the electron, and the quantum numbers $nl$ are now dropped from the radial function $\mathcal{R}$ for simplicity.
When the electron is on core B, thus forming a negative ion, and the system is in the ionic configuration $i$, the appropriate asymptotic form is used
\begin{equation}
\mathcal{R}_i(r) = N_i \frac{2\gamma_i}{\pi} k_{l_i}(\gamma_i r)
\label{eq:wf_ion_gen}
\end{equation}
where $-\gamma_i^2/2$ is the binding energy of the electron, and thus $E_\mathrm{B}^- = \gamma_i^2/2$ is the electron affinity of the negative ion, $k_l(x)$ is the modified spherical Bessel function (e.g. \cite{OlverNISThandbookmathematical2010}), and $l_i$ the orbital angular momentum quantum number.  For the specific case of $l_i=0$,
\begin{equation}
\mathcal{R}_i(r) = N_i \exp{(-\gamma_i r)} / r.
\label{eq:wf_ion}
\end{equation}

Both wavefunctions contain leading constants, $N_c$ and $N_i$, that must be determined.  In the case of the $N_c$, this can be estimated from the Coulomb wavefunction normalisation constant from quantum defect theory \cite{HartreeWaveMechanicsAtom1928, Bates1949, Seaton1958a} (see, e.g.,  eqn. 19 of Ref.~\cite{Janev1972}), which can be expected to give accurate results for sufficiently excited states with diffuse wavefunctions well described by quantum defect theory.  The leading constant factor for the negative ion function is, however, more problematic to determine.  Due to the $1/r$ factor, the normalisation is dominated by contributions at short distances where the asymptotic form is not valid and overestimates the wavefunction, and the value given by normalisation $N_i=\sqrt{2 \gamma_i}$ (see Ref.~\cite{SmirnovAtomicstructureresonant2001}) cannot be expected to give an accurate representation of the asymptotic behaviour.    $N_i$ should therefore be determined by comparison with detailed atomic structure calculations.  One expects $N_i>\sqrt{2 \gamma_i}$, and this has sometimes led to the use of a cutoff value $r_0$, below which the wavefunction is zero and can be chosen to give correct normalisation e.g. Ref.~\cite{SmirnovAtomicstructureresonant2001,barklem_excitation_2016}.

Using these wavefunctions, Refs.~\cite{JanevNonadiabaticTransitionsIonic1976, Janev1976} give the expression for $\Delta(R_x)$ as
\begin{multline}
\Delta(R_x) = \frac{\delta_{m_i m_c} N_i \mathcal{R}_c(R_x)}{(2 \gamma_i R_x)^{|m_c|}|m_c|!}  \\ 
\times \left[  \frac{(2l_c+1)(2l_i+1)(l_c+|m_c|)!(l_i+|m_i|)!}{(l_c-|m_c|)!(l_i-|m_i|)!} \right]^{1/2},
\label{eq:lhj}
\end{multline}
which for $l_i=0$, as in the case of H$^-$ ($^1$S), reduces to the rather simple expression
\begin{equation}
\Delta(R_x) =  N_i \mathcal{R}_c(R_x)  \times (2l_c+1)^{1/2}.
\label{eq:lhj_0}
\end{equation}
The modification due to angular momentum coupling to core electrons in complex systems, including those with equivalent electrons, is given in Refs.~\cite{JanevNonadiabaticTransitionsIonic1976, Janev1976, Chibisov1988}.  The major advantage of this approach is that, as can be seen, it reduces to an analytic expression, and does not require the evaluation of any matrix elements of the Hamiltonian.

\subsection{LCAO method}
\label{sect:lcao}

The LCAO approach can be used to calculate $\Delta(R_x)$ through calculation of the matrix elements of the Hamiltonian using the same or similar atomic wavefunctions for the active electron.   This has been done in Refs.~\cite{Grice1974, Adelman1977, Anstee1992, barklem_excitation_2016} for the case involving H$^-$, considering two electrons.  The main differences between these different descriptions are the treatment of the two-electron wavefunctions, leading to slightly different expressions, but they are equivalent asymptotically (at large separations $R$). For the purposes of our discussion, a much simpler one-electron version is preferable, again equivalent to the above formulations at large $R$.  

Considering the model and coordinate system given in Fig.~\ref{fig:system}, ignoring any interactions with neutral atom B, the Hamiltonian is $H=-1/r_A$, and we have the rather simple expression in terms of atomic wavefunctions
\begin{equation}
H_{ii} = \langle \varphi_i | - \frac{1}{r_A}  | \varphi_i \rangle,
\end{equation}
which for large $R$ we have
\begin{equation}
\langle \varphi_i | \frac{1}{r_A}  | \varphi_i \rangle \approx \frac{1}{R}, 
\end{equation}
and similarly 
\begin{equation}
H_{ic} = \langle \varphi_i |  -\frac{1}{r_A}  | \varphi_c \rangle. 
\end{equation}
Using these results in eqn.~\ref{eq:deltaU}, we obtain
\begin{eqnarray}
\Delta(R_x) & \approx &  2 \left( -\frac{1}{R} \langle \varphi_i | \varphi_c \rangle - \langle \varphi_i |  -\frac{1}{r_A}  | \varphi_c \rangle \right) \bigg|_{R=R_x} \nonumber \\
& \approx & 2  \langle \varphi_i | \frac{1}{r_A}   - \frac{1}{R}  | \varphi_c \rangle \bigg|_{R=R_x}, 
\label{eq:lcao}
\end{eqnarray}
where we now define
\begin{equation}
T_{ic}(R)  \equiv  2  \langle \varphi_i | \frac{1}{r_A}   - \frac{1}{R}  | \varphi_c \rangle,
\end{equation}
the ``post-interaction operator'' $T$ of charge transfer theory also found in Refs.~\cite{Grice1974, Adelman1977}.  The same result is found through a different derivation in \S~10.3 of Ref.~\cite{SmirnovPhysicsatomsions2003}.

Analytic expressions for $T_{ic}(R)$ can be obtained adopting the asymptotic forms of the atomic wavefunctions given above, though for any given $l_c$ and $l_i$, they are significantly more complex than those arising from the surface integral method for $\Delta(R_x)$ given above.  Expressions both for $T_{ic}(R)$ and $S_{ic}(R)$ for the $l_c=l_i=0$ case are given in the Appendix.

\subsection{H$^-$ asymptotic wavefunction}

The general theory of asymptotic wavefunctions has been well expounded, for example in \S~2.1 of Ref.~\cite{PatilAsymptoticmethodsquantum2013}, as well as Ref.~\cite{PatilAsymptoticbehaviourwavefunctions1995}.
The derivation of the asymptotic wavefunction for H$^-$ is particularly simple, and so it is instructive for the following discussion to present the basic equations; see also e.g. Refs.~\cite{Anstee1992, PatilAsymptoticbehaviourwavefunctions1995, SmirnovPhysicsatomsions2003} for details.  The two-electron non-relativistic Hamiltonian, in atomic units, is
\begin{equation}
H = - \frac{1}{2}\nabla_1^2 - \frac{1}{2}\nabla_2^2  - \frac{1}{r_1} - \frac{1}{r_2} + \frac{1}{r_{12}},  
\end{equation}
where $r_1$ and $r_2$ are distances of the electrons with labels 1 and 2 to the proton, and $r_{12}$ is the distance between electrons.
Taking the asymptotic case $r_1 \gg r_2$, then $r_{12}\approx r_1$, the asymptotic form then allows separation into terms related to each electron
\begin{equation}
H \approx (- \frac{1}{2}\nabla_1^2) + (- \frac{1}{2}\nabla_2^2  - \frac{1}{r_2}) \equiv H_1 + H_2,  
\end{equation}
such that $H_1$ is the Hamiltonian for the loosely bound distant electron, and $H_2$ is the Hamiltonian for the tightly bound electron close to the proton.  Then the time-independent Schr\"odinger equation becomes 
\begin{equation}
 (H_1 + H_2) \psi = (-E_{\mathrm{H}^-} + E_\mathrm{H}^\mathrm{1s}) \psi
\end{equation}
where $\psi$ is the electronic wavefunction, $E_\mathrm{H}^\mathrm{1s}$ is the energy of the hydrogen ground state ($-0.5$~au) and $E_{\mathrm{H}^-}$ is the electron affinity of H (a positive value, see below).  Separating the total wavefunction $\psi$ into spatial and spin functions $\psi = \varphi(\vec{r_1}, \vec{r_2}) \chi_{S=0}$, where $\chi_{S=0}$ is the singlet spin function, then the asymptotic spatial wavefunction becomes 
\begin{equation}
 \varphi^\mathrm{asymp}(\vec{r_1}, \vec{r_2}) = \varphi_{LR} (\vec{r_1}) \varphi_{1s}^\mathrm{H} (\vec{r_2})
 \label{eqn:asymp}
\end{equation}
where $H_2 \varphi_{1s}^\mathrm{H} (\vec{r_2}) = E_\mathrm{H}^\mathrm{1s} \varphi_{1s}^\mathrm{H} (\vec{r_2})$ and thus $\varphi_{1s}^\mathrm{H} (\vec{r}) = \exp{(-r)}/\sqrt{\pi}$ is the hydrogen ground state function, and the long range (LR) electron function is the solution of
\begin{equation}
H_1 \varphi_{LR} (\vec{r_1}) = (- \frac{1}{2}\nabla_1^2) \varphi_{LR} (\vec{r_1}) = -E_{\mathrm{H}^-} \varphi_{LR} (\vec{r_1}).
\end{equation}
The spherically symmetric ($l=0$) solution is
\begin{equation}
\varphi_{LR} (\vec{r}) = A \exp{(-\gamma r)}/r   \qquad  \gamma r \gg 1
\label{eqn:lr}
\end{equation}
where $\gamma = \sqrt{2 E_{\mathrm{H}^-}}$, and $A$ is an arbitrary constant related to $N_i$ above in eqn.~\ref{eq:wf_ion}, which we need to determine.  The electron affinity of H from modern measurements is $E_{\mathrm{H}^-} = 0.754195(19)$~eV or 0.0277162 atomic units \cite{LykkeThresholdphotodetachment1991}, which gives $\gamma = 0.235441$ in atomic units.  The validity condition $\gamma r \gg 1$ thus requires $r \gg 1/\gamma = 4.26$~au.  As will be shown below, $A = N_i \frac{1}{\sqrt{4\pi}}$ and correct normalisation would imply $N_i=\sqrt{2\gamma}\approx0.686$ and $A \approx 0.194$.

Unfortunately a range of different definitions and notations for the asymptotic function and the leading constant have been used in various sources.  Part of this difference arises often from separation into radial and angular components, for this case ($l=0$)
\begin{equation}
\varphi_{LR} (\vec{r}) = \mathcal{R}(r) Y_{00}(\theta, \phi) = \mathcal{R}(r) \frac{1}{\sqrt{4\pi}}.
\end{equation}  
In early papers by Smirnov \cite{smirnov_formation_1965}, the following definition was used
\begin{equation}
\mathcal{R}(r) = A_\mathrm{S65} \sqrt{2 \gamma} \exp{(-\gamma r)}/r
\end{equation}
where the $\sqrt{2 \gamma}$ arises since the function is correctly normalised if $A_\mathrm{S65}=1$ (see \S~\ref{sect:surface}).  In Ref.~\cite{smirnov_formation_1965}, the value $A_\mathrm{S65}^2 = 2.65$ was derived, which means $A_\mathrm{S65} = 1.63$.  This implies $A = A_\mathrm{S65} \sqrt{2 \gamma} \frac{1}{\sqrt{4\pi}} = 0.316$.  Later work by Smirnov and others \cite{smirnov_asymptotic_1973, SmirnovAtomicstructureresonant2001, SmirnovPhysicsatomsions2003, Janev1972, Janev1978} have adopted the definition 
\begin{equation}
\mathcal{R}(r) = A_\mathrm{S} \exp{(-\gamma r)}/r = B_\mathrm{S} \sqrt{2 \gamma}  \exp{(-\gamma r)}/r
\end{equation}
such that $B$ in Ref.~\cite{SmirnovPhysicsatomsions2003}, here denoted $B_S$, is equal to $A_\mathrm{S65}$. Note, $A_S = N_i$ and is thus the value required for use in the equations of Janev and Salin \cite{Janev1972,Janev1976}, that is, equations \ref{eq:lhj} and \ref{eq:lhj_0}.   Smirnov \cite{smirnov_asymptotic_1973, SmirnovAtomicstructureresonant2001, SmirnovPhysicsatomsions2003} has derived $A_S$ from the simple Chandrasekhar 2- and 3-parameter wavefunctions \cite{chandrasekhar_remarks_1944}, finding a value of $A_\mathrm{S} = 1.13 \pm 0.06$, and thus $B_\mathrm{S} = A_\mathrm{S65} = 1.64 \pm 0.08$.  A value of $A_S$ in this range has been used in all applications of the LHJ method to processes involving H$^-$ \cite{Janev1978, Janev2006, Jian-GuoMutualrecombinationslow2006, MiyanoHedberg2014, barklem_excitation_2016, barklem_erratum:_2017}.  This gives $A = A_\mathrm{S} \frac{1}{\sqrt{4\pi}} = 0.319$.

In LCAO applications \cite{Adelman1977, Anstee1992} the value $A$ = 0.223106 has been used, determined from the 203-parameter variational wavefunction of Pekeris \cite{pekeris_ground_1958}, asymptotic values of the wavefunction being presented in Ref.~\cite{Ohmura1960}.  This derivation will be discussed below in \S~\ref{sect:direct}.   In Ref.~\cite{barklem_erratum:_2017} the LCAO method was applied, and the same value is essentially used, but a factor of $\sqrt{2}$ must be applied to adjust for the antisymmetrized version of the asymptotic wavefunction that was employed.  

Clearly, there is a discrepancy between the leading constant of the asymptotic wavefunction derived by Smirnov and used in all LHJ theory applications, $A\approx 0.319$, and the value that has been applied in LCAO calculations, i.e. $A\approx 0.223$.  The two values of $A$ differ by a factor of roughly $\sqrt{2}$.  In the following we repeat both derivations in detail to show the source of the problem.


\subsubsection{Direct matching of the asymptotic wavefunction}
\label{sect:direct}

The most straightforward approach to determining the appropriate constant in the asymptotic wavefunction is through direct matching to a calculated wavefunction for H$^-$.  This is the approach that has been used in the LCAO case, and the method stems from Ref.~\cite{Ohmura1960}.  They define a function equal to the two-electron spatial wavefunction where either $r_1=0$ or $r_2=0$, $\psi(r)$ in their notation.  Here we denote this same function: 
\begin{eqnarray}
 \varphi_0(r) & \equiv &  \varphi(\vec{r_1}=0, |\vec{r_2}|=r) =  \varphi(|\vec{r_1}|=r, \vec{r_2}=0) \nonumber \\ & = & C(r) \exp{(-\gamma r)}/r.
 \label{eq:asymp}
\end{eqnarray}  
such that $C(r)$ is the same as in Ref.~\cite{Ohmura1960}, and converges to a constant value $C(\infty)$ at large $r$ \cite{OhmuraLowEnergyElectronHydrogen1958}.   Equating to the asymptotic wavefunction from above, eqn.~\ref{eqn:asymp}, taking $r_2=0$, we obtain
\begin{eqnarray}
\varphi_0^\mathrm{asymp}(r) & = & \varphi_{LR} (|\vec{r_1}|=r) \varphi_{1s}^H (\vec{r_2}=0) \nonumber \\
 & = &  \varphi_{LR} (r) \frac{1}{\sqrt{\pi}} \nonumber \\
 & = & A \exp{(-\gamma r)}/r \frac{1}{\sqrt{\pi}},  \label{eqn:asymp_zero}
\end{eqnarray}
which means $A = C(\infty)  \sqrt{\pi}$.  In Ref.~\cite{Ohmura1960}, $C(\infty) = 0.125874$ and thus $A=0.223106$ was found by matching points at $r=14$--16 atomic units from the calculated values of $\varphi_0(r)$ extracted from the 203-parameter wavefunction of Ref.~\cite{pekeris_ground_1958} given in their Table 1.   This is the value used in LCAO calculations.  

\begin{figure}
\includegraphics[width=0.49\textwidth,angle=0]{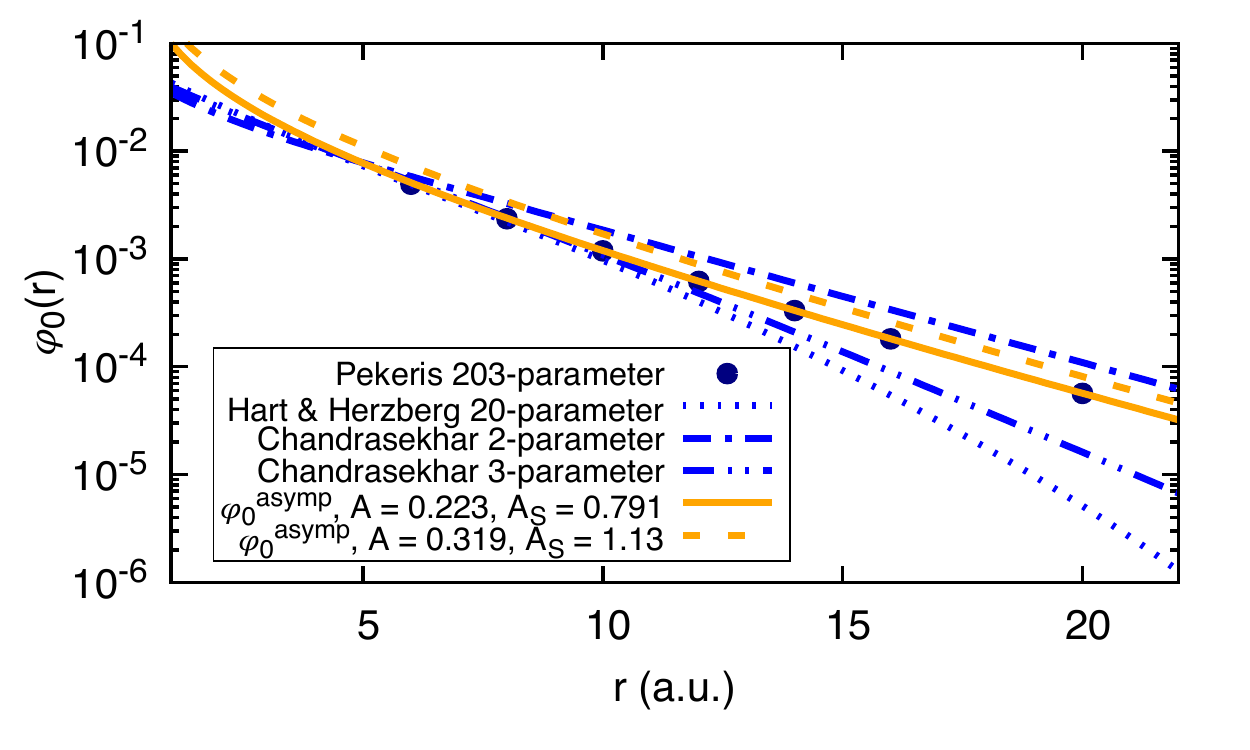}
\caption{
The asymptotic H$^-$ wavefunction $\varphi^\mathrm{asymp}(r)$ as defined in eqn.~\ref{eq:asymp}, extracted from various detailed calculations of the wavefunction $\varphi(\vec{r_1}, \vec{r_2})$: the 203-parameter function from Pekeris \cite{pekeris_ground_1958} calculated in Ref.~\cite{Ohmura1960}, the 19-parameter function of Hart and Herzberg \cite{HartTwentyParameterEigenfunctionsEnergy1957}, the 2- and 3-parameter functions of Chandrasekhar \cite{chandrasekhar_remarks_1944}.  Asymptotic forms (eqn.~\ref{eqn:asymp_zero}) with the two values of $A$ discussed in the text are also shown.  \label{fig:direct}}
\end{figure}

Figure~\ref{fig:direct} compares $\varphi_0(r)$ extracted from various detailed calculations of the wavefunction $\varphi(\vec{r_1}, \vec{r_2})$, with asymptotic forms of the wavefunction $\varphi^\mathrm{asymp}_0(r)$ for the two different values of $A$ discussed above.  It is clearly seen that $A\sim 0.223$ matches all the functions better than the higher value of $A\sim 0.319$.  Note, among detailed calculations, only the Pekeris wavefunction has the expected asymptotic form even at very long range, since the variational form of the wavefunction used ensures correct asymptotic behaviour.  The 2- and 3-parameter wavefunctions derived by Chandrasekhar \cite{chandrasekhar_remarks_1944} are of particular interest as they were used by Smirnov, but also because they are so simple they can be used to exemplify various issues.  The Chandrasekhar wavefunction is:
\begin{multline}
\varphi(\vec{r_1}, \vec{r_2}) =  N (\exp{(-a r_1 - b r_2)} + \exp{(-a r_2 - b r_1)}) \\ \times (1 + c r_{12})
\end{multline}
where for the two parameter function ($c=0$) he finds through variational calculations $a = 1.03925$ and $b = 0.28309$, and it can be found from the normalisation condition that $N = 0.031443$, and for the three parameter function introducing correlation he finds $a = 1.07478$, $b = 0.47758$, $c=0.31214$, and it can be found that $N = 0.031226$.  That $a\approx 1$ in both cases shows the expected result of an unscreened hydrogen $1s$-like function for the electron close to the proton, with a more distant long range electron.  Having said that, it is clear that the functions do not have the correct asymptotic behaviour when the distant electron is very far from the nucleus.  The Chandrasekhar 2-parameter function does not have asymptotic behaviour at any distance, and the 3-parameter and Hart and Herzberg 20-parameter function \cite{HartTwentyParameterEigenfunctionsEnergy1957} have only roughly asymptotic behaviour in the region $r\approx 4$--10~au.  This is due to the fact that the asymptotic region has little influence on the state energies and thus the wavefunction determined by variational methods. 

\subsubsection{Electronic density matching method}

\begin{figure}
\includegraphics[width=0.49\textwidth,angle=0]{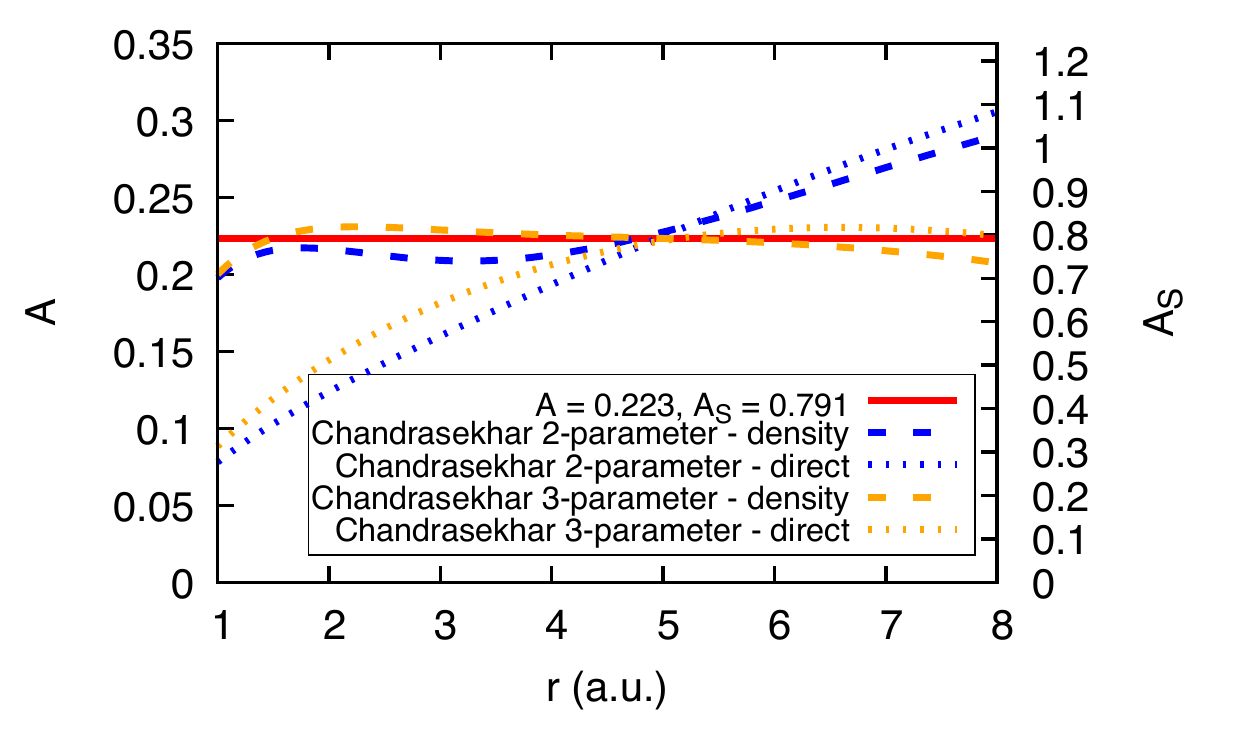}
\caption{
Derived values for $A$ as a function of $r$.  Corresponding values for $A_\mathrm{S}$ are shown on the right axis.  The blue lines show the values derived from Chandrasekhar's 2-parameter wavefunction, and the orange lines from the 3-parameter wavefunction.  In each case, dashed lines are from the electronic density matching method, and dotted lines from the direct matching method.  The horizontal full red line shows the value derived from direct matching to the accurate wavefunction due to Pekeris \cite{pekeris_ground_1958}; see Fig. \ref{fig:direct} and \S~\ref{sect:direct}.  \label{fig:chandra}}
\end{figure}

An alternative approach, and that which has been used by Smirnov \cite{smirnov_asymptotic_1973, SmirnovAtomicstructureresonant2001, SmirnovPhysicsatomsions2003}, is to match the electron density function.  Generally for an $N$-electron system, and ignoring spin, the electronic density is
\begin{equation}
\rho(\vec{r}) = N \int d\vec{r_2} ... \int d\vec{r_N} |\varphi(\vec{r}, \vec{r_2}, ..., \vec{r_N})|^2,
\end{equation}
and thus for H$^-$ and two electrons
\begin{equation}
\rho(\vec{r}) = 2 \int d\vec{r_2} |\varphi(\vec{r}, \vec{r_2})|^2. 
\end{equation}
For the asymptotic form this gives:
\begin{eqnarray}
\rho^\mathrm{asymp} (\vec{r}) & = &  2 \int d\vec{r_2} |\varphi_{LR} (\vec{r}) \varphi_{1s}^H (\vec{r_2})|^2  \nonumber \\
         & = & 2 |\varphi_{LR} (\vec{r})|^2 \nonumber \\
         & = &  2 A^2 \exp{(-2\gamma r)}/r .
\end{eqnarray}
As done by Smirnov, one can derive analytical expressions for the asymptotic constant $A$ or some related version of it, e.g. $A_\mathrm{S}$ or $A_\mathrm{S65}$, as a function of $r$, by equating the asymptotic form with the electronic density calculated from Chandrasekhar's wavefunctions, and the expressions found by Smirnov are given in Refs.~\cite{SmirnovAtomicstructureresonant2001, SmirnovPhysicsatomsions2003}.  Using Mathematica this analysis was repeated here, making use of Hylleraas coordinates for the 3-parameter case \cite{HylleraasNeueBerechnungEnergie1929,PanHylleraasCoordinates2003}, and the results are plotted in Fig.~\ref{fig:chandra}.

From this figure it can be seen that the values of $A$ and $A_S$ as a function of $r$ from electronic density matching of both the 2- and 3-parameter Chandrasekhar wavefunctions are roughly constant and in agreement around $r=2$--5~au, {the region used by Smirnov to derive $A_S$}.  The figure clearly favours a low value $A\approx0.22$ ($A_S \approx 0.79$), rather than a high value $A\approx0.32$ ($A_S \approx 1.13$). Thus, the analysis presented here from electron density matching is roughly in agreement with the value from direct matching, $A=0.223$ ($A_S=0.791$), which is roughly a factor of $1/\sqrt{2}$ lower than found by Smirnov $A\approx0.32$ ($A_S\approx1.13$).  The difference is not precisely a factor of $1/\sqrt{2}$ since different wavefunctions are used, the direct method permitting the easy use of the accurate 203-parameter function of Pekeris.

Note, the criterion for validity of the asymptotic wavefunction, $r \gg 1/\gamma = 4.26$~a.u., is not satisfied across this entire region $r=2$--5~a.u.  Further, it is clear that the Chandrasekhar wavefunctions do not have correct asymptotic behaviour, as is seen from application of the direct method, which gives rapidly varying values of $A$ (see also Fig.~\ref{fig:chandra}).   That the density matching method using these simple wavefunctions gives values that agree reasonably with the values from direct matching of the accurate and asymptotically correct wavefunction of Pekeris, is thus perhaps somewhat surprising.  It may reflect that the density method, by accounting for contributions from both electrons at intermediate distances, reaches asymptotic behaviour at smaller internuclear distances than the direct method where the second electron is fixed at the nucleus.  Contributions from both electrons to the wavefunction are equal, $\varphi_{LR} (\vec{r}) = \varphi^H_{1s} (\vec{r})$, at roughly $r=2.3$~a.u.

\section{Comparison of theoretical calculations and experiments}

Experiments have recently been performed in Louvain on MN of Li$^+$/D$^-$ \cite{LaunoyMutualNeutralizationLi2019} and at the DESIREE facility in Stockholm on Li$^+$/D$^-$ \cite{EklundCryogenicmergedionbeamexperiments2020} and Na$^+$/D$^-$ \cite{EklundFinalstateresolvedmutualneutralization2021}, resolving final states and thus measuring branching fractions for the neutral products. D$^-$ is preferred over H$^-$ in the experiments for practical reasons, but is basically identical to H$^-$ in terms of binding energy (electron affinity) and thus electronic structure.  The different mass leads to trajectory (Coulomb focussing) effects, but is easily accounted for in the dynamical calculations.  In a recent paper \cite{BarklemMutualNeutralizationLi2021}, existing full quantum calculations \cite{Croft1999a, Croft1999, dickinson_initio_1999} and (two-electron) LCAO method calculations \cite{barklem_excitation_2016, barklem_erratum:_2017} were compared with experiment, generally finding good agreement, with the full quantum calculations performing best, in line with expectations.  The reader is referred to Ref.~\cite{BarklemMutualNeutralizationLi2021} for a discussion regarding these comparisons, including the discussion of Coulomb focussing effects caused by H$^-$ versus D$^-$.  

\begin{figure}
\includegraphics[width=0.49\textwidth,angle=0,trim={0 25mm 0 0},clip]{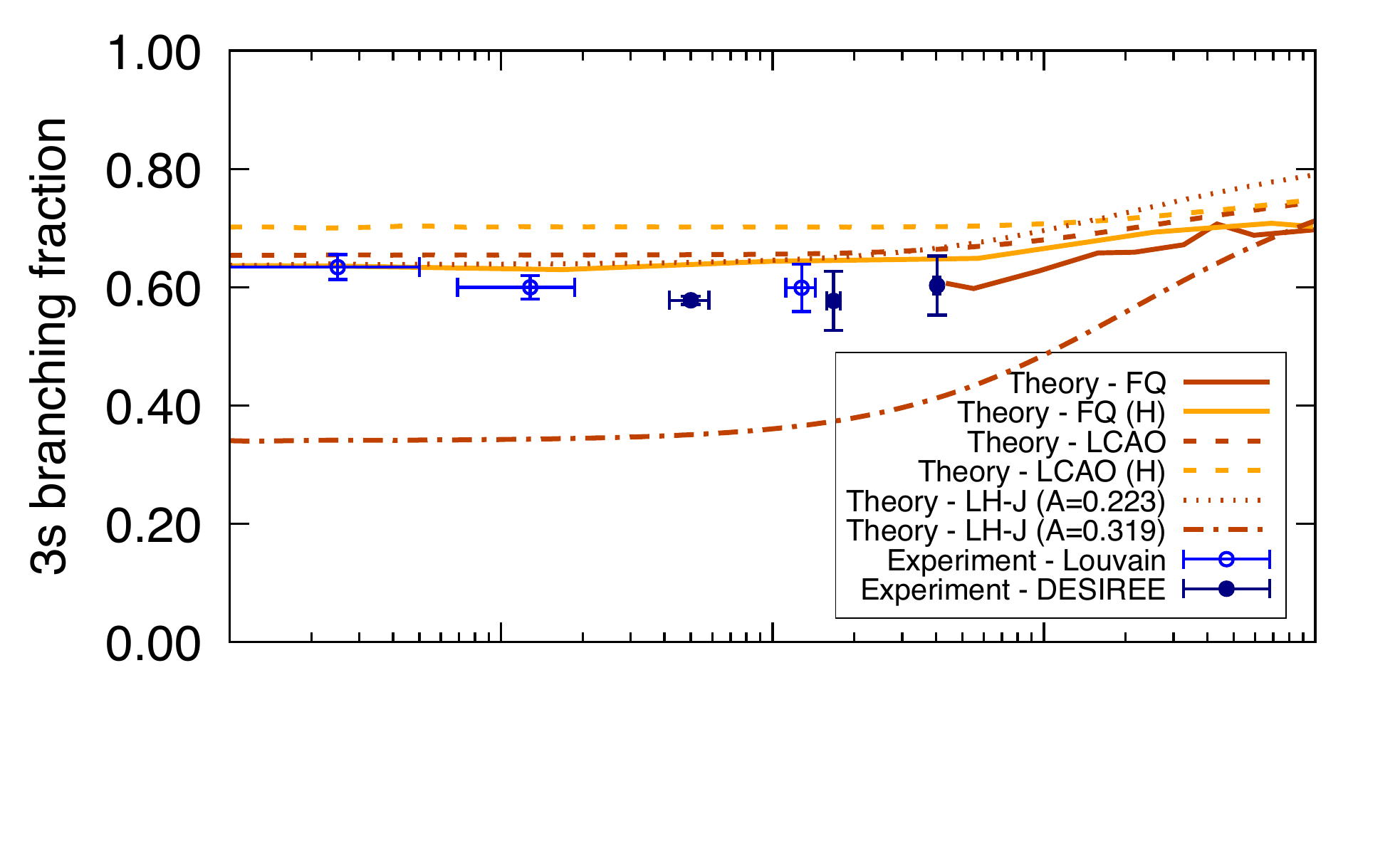}\\
\includegraphics[width=0.49\textwidth,angle=0,trim={0 25mm 0 0},clip]{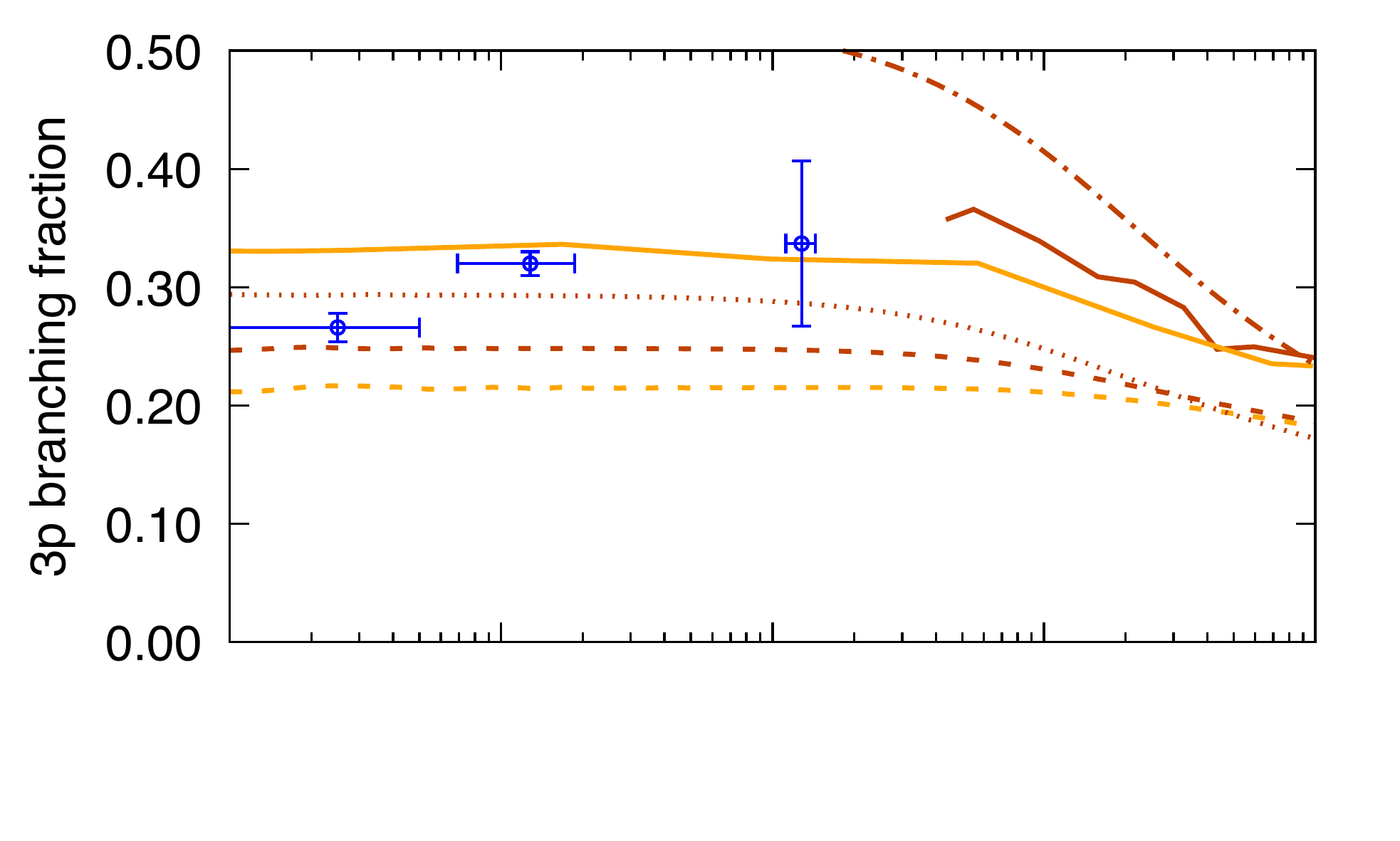}\\
\includegraphics[width=0.49\textwidth,angle=0]{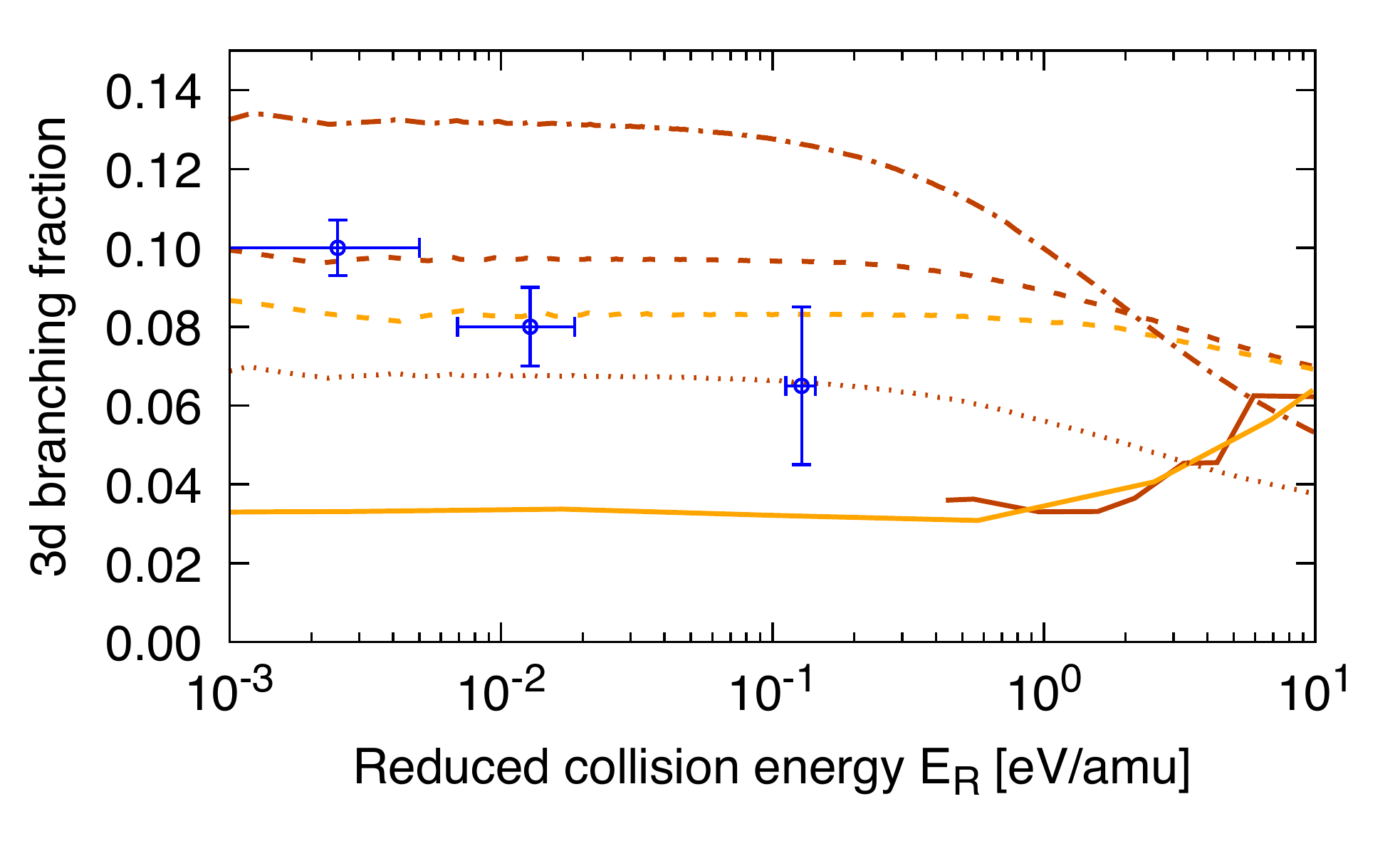}
\caption{ 
Branching fractions for the MN reaction $\mathrm{Li}^+ + \mathrm{D}^- \rightarrow \mathrm{Li}(nl) + \mathrm{D}$ as a function of reduced collision energy.  The $3s$, $3p$ and $3d$ channels are shown in separate panels.  
Experimental results from Louvain \cite{LaunoyMutualNeutralizationLi2019} and from DESIREE \cite{EklundCryogenicmergedionbeamexperiments2020} are shown, with estimated errors ($1\sigma$).  
Theoretical results are shown from full quantum (FQ) calculations, LCAO, and LHJ asymptotic methods with the old (high) value for $A$, and the new (low) value for $A$; see text for further details. Theoretical results are also shown for the case where D$^-$ is replaced by H$^-$, marked by (H) in the legend.
\label{fig:Li_bf}}
\end{figure}

\begin{figure}
\includegraphics[width=0.49\textwidth,angle=0,trim={0 25mm 0 0},clip]{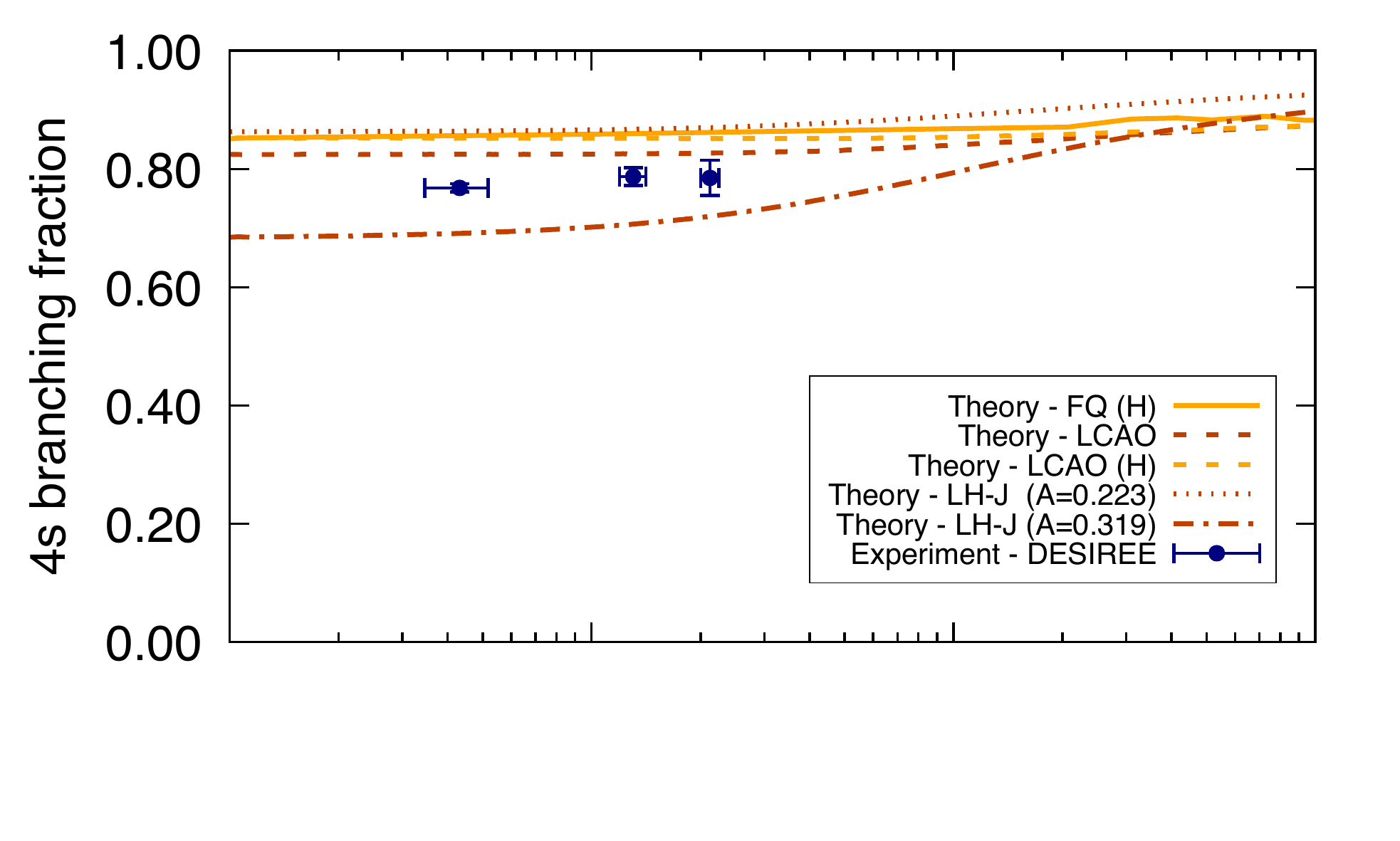}\\
\includegraphics[width=0.49\textwidth,angle=0,trim={0 25mm 0 0},clip]{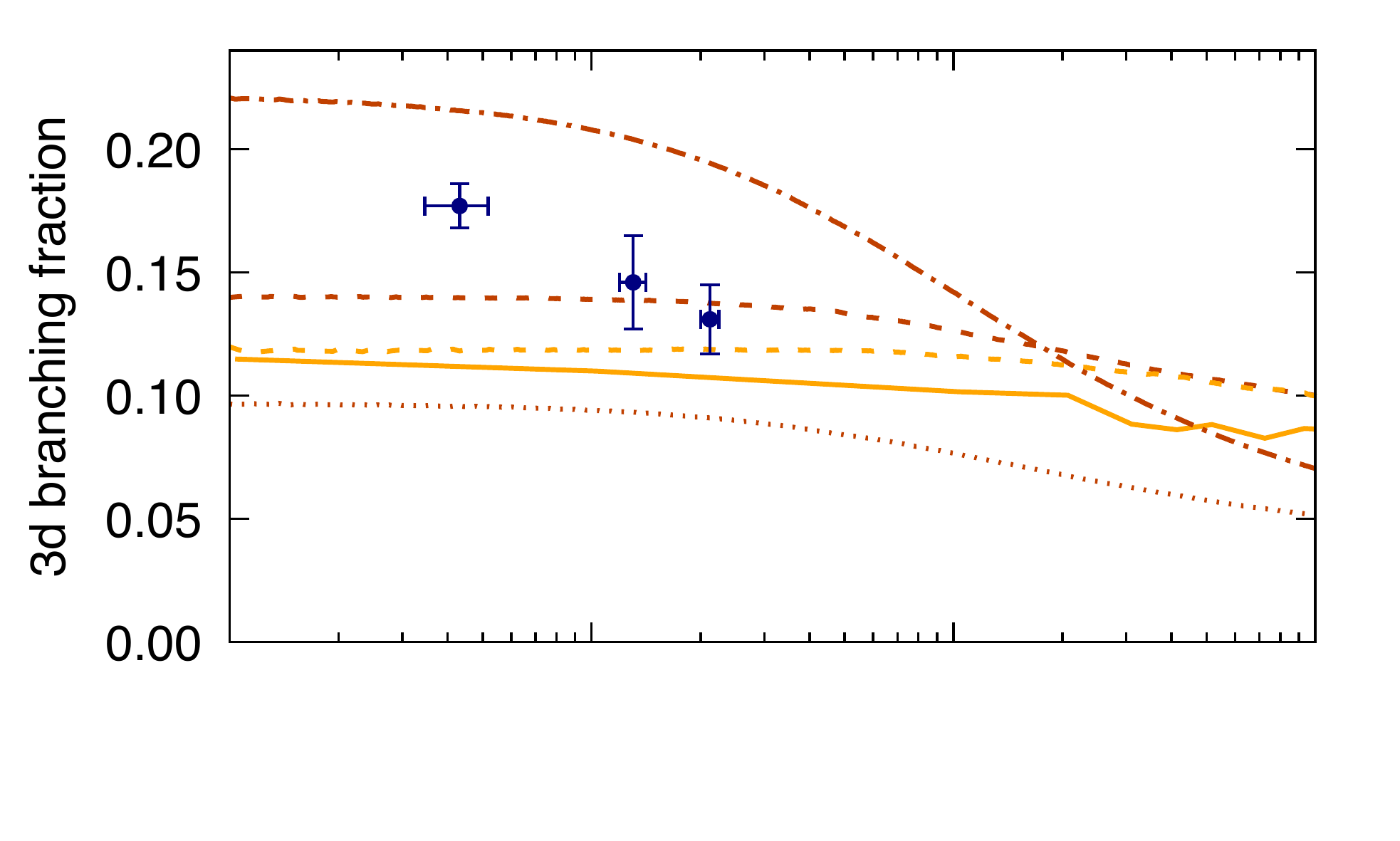}\\
\includegraphics[width=0.49\textwidth,angle=0]{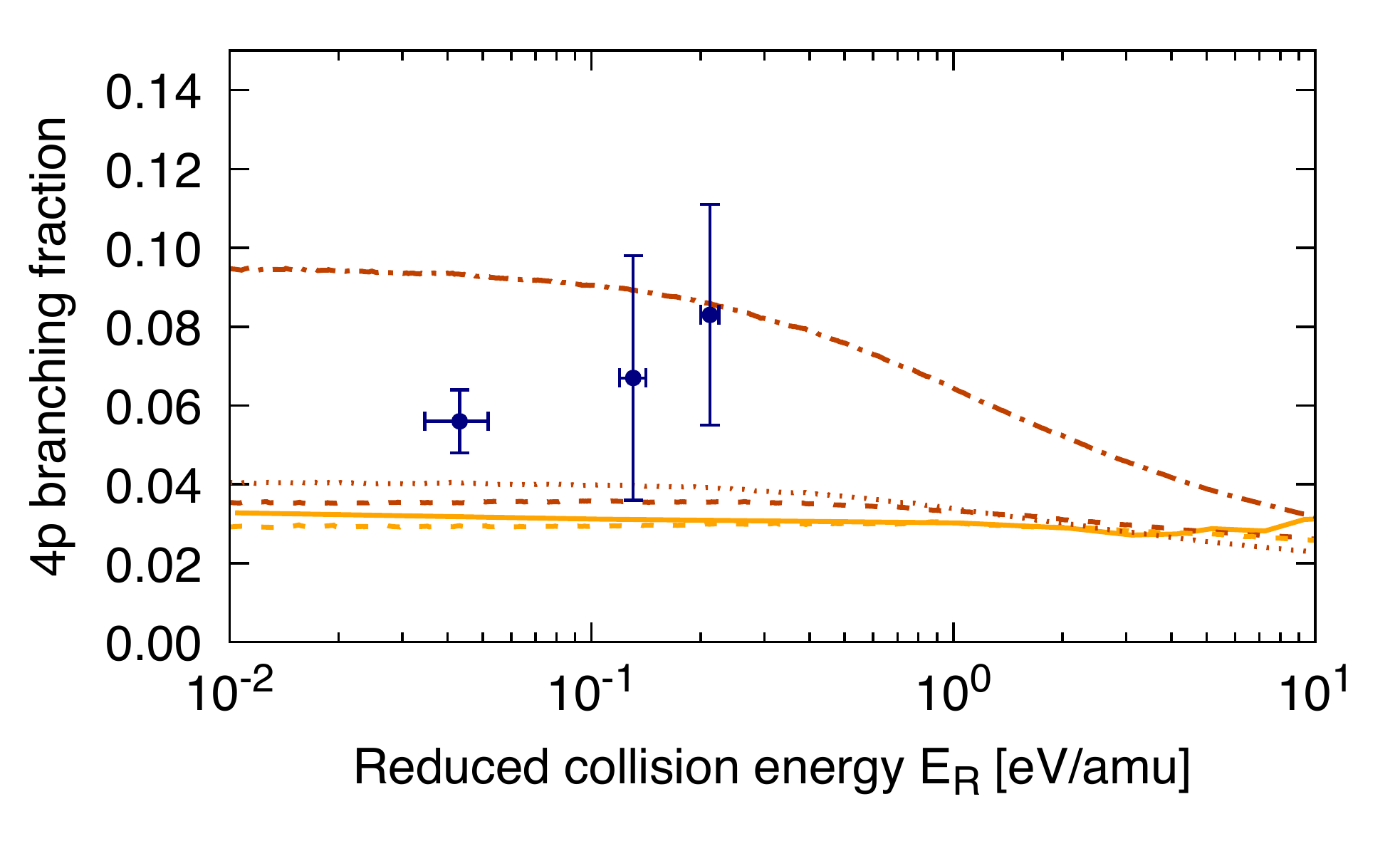}
\caption{ 
Branching fractions for the MN reaction $\mathrm{Na}^+ + \mathrm{D}^- \rightarrow \mathrm{Na}(nl) + \mathrm{D}$ as a function of reduced collision energy.  The $4s$, $3d$ and $4p$ channels are shown in separate panels.  Experimental results from DESIREE \cite{EklundFinalstateresolvedmutualneutralization2021} are shown, with estimated errors ($1\sigma$).  Theoretical results are shown from full quantum (FQ) calculations, LCAO and LHJ asymptotic methods with the old (high) value for $A$, and the new (low) value for $A$; see text for further details.  Theoretical results are also shown for the case where D$^-$ is replaced by H$^-$, marked by (H) in the legend.
\label{fig:Na_bf}}
\end{figure}

Here, these comparisons are supplemented with calculations from the LHJ theory using the two $A$ values discussed, and these calculations use the same code as the LCAO calculations, described in Refs.~\cite{barklem_excitation_2016, barklem_erratum:_2017}.  The results are shown for Li$^+$/D$^-$ in Fig.~\ref{fig:Li_bf}, and for Na$^+$/D$^-$ in Fig.~\ref{fig:Na_bf}.  To aid comparison (see Ref.~\cite{BarklemMutualNeutralizationLi2021}), the plots are presented on the reduced energy scale, which is defined as $E_\mathrm{R} = E_\mathrm{CM}/\mu = \frac{1}{2}v^2$, where $E_\mathrm{CM}$ is the collision energy in the centre-of-mass frame, $\mu$ is the reduced mass of the system, and $v$ is the relative velocity.  It is clear that the comparisons with experiment are significantly improved for both Li$^+$/D$^-$ and Na$^+$/D$^-$ if the low value of $A=0.223$ is adopted, compared to the high value of $A=0.319$. In particular we note that for Li$^+$/D$^-$ (Fig.~\ref{fig:Li_bf}), the high value leads to the prediction of $3p$ being the dominant channel at low energy, rather than $3s$ as clearly seen to a high degree of certainty in experiments.  This discrepancy is resolved for the LHJ theory with the lower value of $A$.  Further the low value of $A$ leads to much better consistency between full quantum, LCAO and LHJ predictions in both cases.  Note also that LHJ with the high value of $A$ leads to variation of the branching fractions at lower collision energies.

Janev and Radulovic \cite{Janev1978} also performed calculations with the LHJ method, for Li$^+$/H$^-$ and Na$^+$/H$^-$, though state-resolved results are only shown for Na$^+$/H$^-$ (Fig.~1 of Ref.~\cite{Janev1978}).  Our results for Na$^+$/H$^-$ partial cross sections are significantly different from theirs; in particular at low energy they find $3p$ to be the second most populated channel, only an order of magnitude lower than $4s$.  This is in strong disagreement with experiment where the $3p$ channel is not observed, and with all other calculations, including our LHJ calculations with the old, high value of $A$, where we find the cross section for $3p$ to be more than five orders of magnitude lower than $4s$; this may suggest a labelling error. In Ref.~\cite{dickinson_initio_1999} substantial differences were also noted compared to full quantum calculations.  This was investigated further by comparing our LHJ theory avoided crossing parameters with those in Table~1 of Ref.~\cite{Janev1978}, in particular $\Delta(R_x)$.  However, for both Li and Na we were unable to reproduce all of their results using the equations and numbers in the paper.  Using their equations, their value for the leading constant in the negative ion radial wavefunction, and their values for $R_x$ we obtain $\Delta(R_x)$ in agreement (within 10\%) for a few crossings (Li $2s$ and $2p$, Na $3p$), but for the majority there are significant differences (up to a factor of 2.5 for Na $4p$), usually our result being larger (only for Li $3d$ do we obtain a roughly 20\% smaller value).  Our LHJ theory values using $A=0.319$ agree well with what we calculate from the equations in  Ref.~\cite{Janev1978}, which are just simpler versions of the more general theory.  As such, detailed comparison of the avoided crossing parameters is not particularly enlightening on the more general problem of interest here, namely the relationship between LHJ theory, LCAO theory, and experiment.

Dochain and Urbain [private communication] have semi-empirically found a similar need for a factor $1/\sqrt{2}$ correction to the couplings in LHJ theory, and in doing so obtain good matches to experimental results for MN of D$^-$ with He$^+$, Li$^+$, Na$^+$, C$^+$, N$^+$, though with some discrepancies for O$^+$.   

Finally, it is worth noting that calculations with the Smirnov formulation of the surface integral method \cite{smirnov_formation_1965}, taking the surface of integration as the midplane between the two atoms, while improved with the lower value of $A$, still perform substantially worse than the Janev and Salin formulation; see also Ref.~\cite{MiyanoHedberg2014}.

\section{Discussion}

It has been claimed in Ref.~\cite{JanevNonadiabaticTransitionsIonic1976}, that the LCAO method is ``essentially incorrect for the problem of calculation of $\Delta(R)$'', with reference to the work of Herring \cite{Herring1962}.  It is, however, unclear that the criticisms of the LCAO method for the cases discussed by Herring, the exchange interaction \footnote{Exchange interaction is used in two senses here, in H$_2$ the usual interaction due to antisymmetrization with respect to exchange of electron labels, and in H$_2^+$, due to the degeneracy with respect to exchange of position of the electron from one nucleus to the other.} in H$_2^+$ and H$_2$, apply also to ionic-covalent interactions.  First, the problem of the $1/r_{12}$ term described for the exchange interaction in H$_2$ is not relevant, as it does not enter the one-electron LCAO method (see S~\ref{sect:lcao}) and in the two-electron methods only enters empirically via the H$^-$ electron affinity and wavefunction (e.g. eqn 10 of Ref.~\cite{barklem_excitation_2016}).  Second, in both cases, a basis with a single tightly bound $1s$ wavefunction is used in the LCAO description.  In the case studied in detail by Herring of the exchange splitting in H$_2^+$, the surface integral approach permits a modification factor which increases the value of the wavefunction between the two nuclei, capturing departures of the molecular wavefunction from the atomic ones, which cannot be modelled in a LCAO basis with a single fixed orbital.  In contrast, the LCAO method for ionic-covalent interactions employs two wavefunctions that capture the main components of the molecular wavefunction at internuclear distances corresponding to the crossing.

\begin{figure}
\begin{tikzpicture}
\pgftext{%
\includegraphics[width=0.45\textwidth,angle=0]{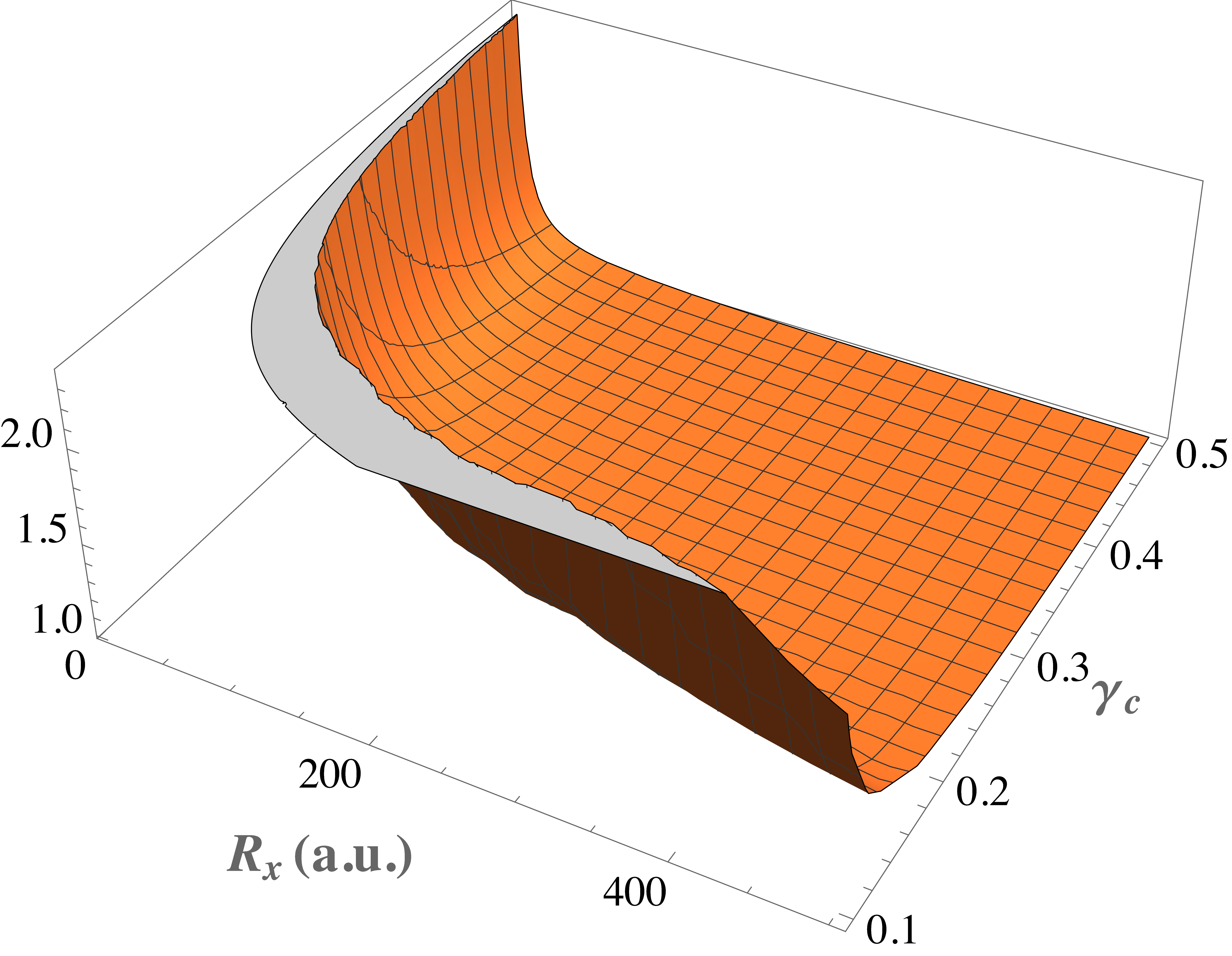}
}%
\node at (4,-.1) {\scriptsize $(n^*=2)$};
\node at (2.0,-3.25) {\scriptsize $(n^*=10)$};
\end{tikzpicture}
\caption{
3D mesh plot of the ratio $\Delta_\mathrm{LCAO}/\Delta_\mathrm{LHJ}$ as a function of $\gamma_c$ and $R_x$, for $l_c=l_i=0$ analytic expressions obtained as described in the text.  The ratio tends to 1 for $R_x \gg \gamma_c^{-2}$; see text.  The plot ranges from $n^*=2$ ($\gamma_c=0.5$) to $n^*=10$ ($\gamma_c=0.1$). 
\label{fig:contour}}
\end{figure}

More quantitatively, in Ref.~\cite{JanevNonadiabaticTransitionsIonic1976}, it is claimed that there is a discrepancy between the asymptotic behaviour of the interactions in the LHJ and LCAO formulations.  For LHJ the interaction has the asymptotic behaviour
\begin{equation}
\Delta_\mathrm{LHJ}(R_x) \propto \exp(-\gamma_c R_x),
\end{equation}
at large $R_x$, see eqns.~\ref{eq:lhj} and ~\ref{eq:wf_cov}, while for LCAO at large $R$, Ref.~\cite{JanevNonadiabaticTransitionsIonic1976} claims the asymptotic behaviour is  
\begin{equation}
\Delta_\mathrm{LCAO}(R_x) \propto \exp(-(\gamma_i + \gamma_c) R_x);
\end{equation}
however, this relationship is asserted without proof, and misses the important fact that $\gamma_i$ and $\gamma_c$ are related at the crossing point $R_x$.  Assuming a pure Coulomb interaction for the ionic potential, and a null potential for the covalent state, the crossing point is given by 
\begin{equation}
\frac{1}{R_x} = \frac{\gamma_c^2}{2} -  \frac{\gamma_i^2}{2}.
\label{eq:rx}  
\end{equation}
Using the one-electron LCAO expression for $T_{ic}(R)$ for $l_c=l_i=0$ given in the Appendix and making the substitution $\gamma_i \rightarrow \sqrt{\gamma_c^2/2 - 2/R_x}$, an analytic expression for $\Delta_\mathrm{LCAO}(R_x) = T_{ic}(R = R_x)$ can be obtained.
This results in a complicated expression, which is no longer dependent on $\gamma_i$,  and which can be shown to have the same behaviour as the LHJ theory for large $R_x$;  i.e., for $l_c=0$
\begin{equation}
\Delta_\mathrm{LHJ}(R_x) =  N_i \mathcal{R}_c(R_x).
\end{equation}
This agreement is most convincingly shown by numerical results for the ratio $\Delta_\mathrm{LCAO}/\Delta_\mathrm{LHJ}$, which is dependent only on $\gamma_c$ and $R_x$, shown in Fig.~\ref{fig:contour}.  The condition for validity of the LHJ method is $R_x\gg \gamma_i^{-2}$ (Ref.~\cite{Janev1972}), which, noting that for large $R_x$ we have $\gamma_i\approx \gamma_c-1/{(\gamma_c R_x)}$, is roughly equivalent to $R_x\gg \gamma_c^{-2}$.  The ratio is seen to be exactly one in the region of validity of the LHJ theory.  Note the effective principle quantum number of the neutral state on A is $n^*=1/\gamma_c$, thus the validity criterion can also be written $R_x \gg {n^*}^{2}$.  Outside the region of validity for the LHJ theory, at crossings at shorter internuclear distances, the LCAO theory gives larger interactions, and the disagreement increases very rapidly.


In summary, previous claims that LCAO theory is incorrect for ionic-covalent interactions at large internuclear distance seem to be unfounded, and correction of the leading constant for the H$^-$ asymptotic wavefunction in the LHJ theory bring the two theories into reasonable quantitative agreement for the two cases of MN of Li$^+$ and Na$^+$ with H$^-$ and D$^-$.  It has also been shown that the LCAO and LHJ theory have the same behaviour for large $R_x$.  The LCAO theory has only been developed for cases involving H$^-$ and D$^-$, treating two electrons to varying degrees of complexity.  The surface integral method, LHJ, is however easily applied to any given situation if the asymptotic wavefunction for the negative ion is known.  The results shown here suggest that if correct wavefunctions for the active electron on the negative ion can be obtained, LHJ theory can be used to obtain reasonable estimates for MN processes occurring at large internuclear distance in practically any system, perhaps even those involving molecules, and thus is of potentially great utility for situations such as astrophysical modelling, where often a large number of rate estimates are needed, and completeness can be more important than accuracy.  It would be useful also to further develop and investigate the utility of the one-electron LCAO method as outlined in this paper, enabling application to complex atoms beyond H$^-$ and D$^-$.

\begin{acknowledgments}
I thank Xavier Urbain, Arnaud Dochain, Jon Grumer, Anish Amarsi, and Gustav Eklund for useful discussions, sharing preliminary results, and comments on the manuscript. The work relies on experimental results obtained at the Swedish National Infrastructure, DESIREE (Swedish Research Council Contract No. 2017-00621), originally published elsewhere, and I thank the DESIREE group at Stockholm University for collaboration on the experiments.  This work is part of the project “Probing charge- and mass- transfer reactions on the atomic level”, supported by the Knut and Alice Wallenberg Foundation (2018.0028), and supported by the Swedish Research Council through individual project grants with contract Nos. 2016-03765 and 2020-03404.    
\end{acknowledgments}

\bibliography{MyLibrary}

\begin{thebibliography}{57}%
\makeatletter
\providecommand \@ifxundefined [1]{%
 \@ifx{#1\undefined}
}%
\providecommand \@ifnum [1]{%
 \ifnum #1\expandafter \@firstoftwo
 \else \expandafter \@secondoftwo
 \fi
}%
\providecommand \@ifx [1]{%
 \ifx #1\expandafter \@firstoftwo
 \else \expandafter \@secondoftwo
 \fi
}%
\providecommand \natexlab [1]{#1}%
\providecommand \enquote  [1]{``#1''}%
\providecommand \bibnamefont  [1]{#1}%
\providecommand \bibfnamefont [1]{#1}%
\providecommand \citenamefont [1]{#1}%
\providecommand \href@noop [0]{\@secondoftwo}%
\providecommand \href [0]{\begingroup \@sanitize@url \@href}%
\providecommand \@href[1]{\@@startlink{#1}\@@href}%
\providecommand \@@href[1]{\endgroup#1\@@endlink}%
\providecommand \@sanitize@url [0]{\catcode `\\12\catcode `\$12\catcode
  `\&12\catcode `\#12\catcode `\^12\catcode `\_12\catcode `\%12\relax}%
\providecommand \@@startlink[1]{}%
\providecommand \@@endlink[0]{}%
\providecommand \url  [0]{\begingroup\@sanitize@url \@url }%
\providecommand \@url [1]{\endgroup\@href {#1}{\urlprefix }}%
\providecommand \urlprefix  [0]{URL }%
\providecommand \Eprint [0]{\href }%
\providecommand \doibase [0]{https://doi.org/}%
\providecommand \selectlanguage [0]{\@gobble}%
\providecommand \bibinfo  [0]{\@secondoftwo}%
\providecommand \bibfield  [0]{\@secondoftwo}%
\providecommand \translation [1]{[#1]}%
\providecommand \BibitemOpen [0]{}%
\providecommand \bibitemStop [0]{}%
\providecommand \bibitemNoStop [0]{.\EOS\space}%
\providecommand \EOS [0]{\spacefactor3000\relax}%
\providecommand \BibitemShut  [1]{\csname bibitem#1\endcsname}%
\let\auto@bib@innerbib\@empty
\bibitem [{\citenamefont {Launoy}\ \emph {et~al.}(2019)\citenamefont {Launoy},
  \citenamefont {Loreau}, \citenamefont {Dochain}, \citenamefont {Li{\'e}vin},
  \citenamefont {Vaeck},\ and\ \citenamefont
  {Urbain}}]{LaunoyMutualNeutralizationLi2019}%
  \BibitemOpen
  \bibfield  {author} {\bibinfo {author} {\bibfnamefont {T.}~\bibnamefont
  {Launoy}}, \bibinfo {author} {\bibfnamefont {J.}~\bibnamefont {Loreau}},
  \bibinfo {author} {\bibfnamefont {A.}~\bibnamefont {Dochain}}, \bibinfo
  {author} {\bibfnamefont {J.}~\bibnamefont {Li{\'e}vin}}, \bibinfo {author}
  {\bibfnamefont {N.}~\bibnamefont {Vaeck}},\ and\ \bibinfo {author}
  {\bibfnamefont {X.}~\bibnamefont {Urbain}},\ }\bibfield  {title} {\bibinfo
  {title} {Mutual {{Neutralization}} in {{Li}}{\textsuperscript{+}} +
  {{D}}{\textsuperscript{-}} {{Collisions}}: {{A Combined Experimental}} and
  {{Theoretical Study}}},\ }\href {https://doi.org/10.3847/1538-4357/ab3346}
  {\bibfield  {journal} {\bibinfo  {journal} {ApJ}\ }\textbf {\bibinfo {volume}
  {883}},\ \bibinfo {pages} {85} (\bibinfo {year} {2019})}\BibitemShut
  {NoStop}%
\bibitem [{\citenamefont {Eklund}\ \emph {et~al.}(2020)\citenamefont {Eklund},
  \citenamefont {Grumer}, \citenamefont {Ros{\'e}n}, \citenamefont {Ji},
  \citenamefont {Punnakayathil}, \citenamefont {K{\"a}llberg}, \citenamefont
  {Simonsson}, \citenamefont {Thomas}, \citenamefont {Stockett}, \citenamefont
  {Reinhed}, \citenamefont {L{\"o}fgren}, \citenamefont {Bj{\"o}rkhage},
  \citenamefont {Blom}, \citenamefont {Barklem}, \citenamefont {Cederquist},
  \citenamefont {Zettergren},\ and\ \citenamefont
  {Schmidt}}]{EklundCryogenicmergedionbeamexperiments2020}%
  \BibitemOpen
  \bibfield  {author} {\bibinfo {author} {\bibfnamefont {G.}~\bibnamefont
  {Eklund}}, \bibinfo {author} {\bibfnamefont {J.}~\bibnamefont {Grumer}},
  \bibinfo {author} {\bibfnamefont {S.}~\bibnamefont {Ros{\'e}n}}, \bibinfo
  {author} {\bibfnamefont {M.}~\bibnamefont {Ji}}, \bibinfo {author}
  {\bibfnamefont {N.}~\bibnamefont {Punnakayathil}}, \bibinfo {author}
  {\bibfnamefont {A.}~\bibnamefont {K{\"a}llberg}}, \bibinfo {author}
  {\bibfnamefont {A.}~\bibnamefont {Simonsson}}, \bibinfo {author}
  {\bibfnamefont {R.~D.}\ \bibnamefont {Thomas}}, \bibinfo {author}
  {\bibfnamefont {M.~H.}\ \bibnamefont {Stockett}}, \bibinfo {author}
  {\bibfnamefont {P.}~\bibnamefont {Reinhed}}, \bibinfo {author} {\bibfnamefont
  {P.}~\bibnamefont {L{\"o}fgren}}, \bibinfo {author} {\bibfnamefont
  {M.}~\bibnamefont {Bj{\"o}rkhage}}, \bibinfo {author} {\bibfnamefont
  {M.}~\bibnamefont {Blom}}, \bibinfo {author} {\bibfnamefont {P.~S.}\
  \bibnamefont {Barklem}}, \bibinfo {author} {\bibfnamefont {H.}~\bibnamefont
  {Cederquist}}, \bibinfo {author} {\bibfnamefont {H.}~\bibnamefont
  {Zettergren}},\ and\ \bibinfo {author} {\bibfnamefont {H.~T.}\ \bibnamefont
  {Schmidt}},\ }\bibfield  {title} {\bibinfo {title} {Cryogenic merged-ion-beam
  experiments in {{DESIREE}}: {{Final}}-state-resolved mutual neutralization of
  {{Li}}{\textsuperscript{+}} and {{D}}{\textsuperscript{-}}},\ }\href
  {https://doi.org/10.1103/PhysRevA.102.012823} {\bibfield  {journal} {\bibinfo
   {journal} {Phys. Rev. A}\ }\textbf {\bibinfo {volume} {102}},\ \bibinfo
  {pages} {012823} (\bibinfo {year} {2020})}\BibitemShut {NoStop}%
\bibitem [{\citenamefont {Eklund}\ \emph {et~al.}(2021)\citenamefont {Eklund},
  \citenamefont {Grumer}, \citenamefont {Barklem}, \citenamefont {Ros{\'e}n},
  \citenamefont {Ji}, \citenamefont {Simonsson}, \citenamefont {Thomas},
  \citenamefont {Cederquist}, \citenamefont {Zettergren},\ and\ \citenamefont
  {Schmidt}}]{EklundFinalstateresolvedmutualneutralization2021}%
  \BibitemOpen
  \bibfield  {author} {\bibinfo {author} {\bibfnamefont {G.}~\bibnamefont
  {Eklund}}, \bibinfo {author} {\bibfnamefont {J.}~\bibnamefont {Grumer}},
  \bibinfo {author} {\bibfnamefont {P.~S.}\ \bibnamefont {Barklem}}, \bibinfo
  {author} {\bibfnamefont {S.}~\bibnamefont {Ros{\'e}n}}, \bibinfo {author}
  {\bibfnamefont {M.}~\bibnamefont {Ji}}, \bibinfo {author} {\bibfnamefont
  {A.}~\bibnamefont {Simonsson}}, \bibinfo {author} {\bibfnamefont {R.~D.}\
  \bibnamefont {Thomas}}, \bibinfo {author} {\bibfnamefont {H.}~\bibnamefont
  {Cederquist}}, \bibinfo {author} {\bibfnamefont {H.}~\bibnamefont
  {Zettergren}},\ and\ \bibinfo {author} {\bibfnamefont {H.~T.}\ \bibnamefont
  {Schmidt}},\ }\bibfield  {title} {\bibinfo {title} {Final-state-resolved
  mutual neutralization of {{Na}}{\textsuperscript{+}} and
  {{D}}{\textsuperscript{-}}},\ }\href
  {https://doi.org/10.1103/PhysRevA.103.032814} {\bibfield  {journal} {\bibinfo
   {journal} {Phys. Rev. A}\ }\textbf {\bibinfo {volume} {103}},\ \bibinfo
  {pages} {032814} (\bibinfo {year} {2021})}\BibitemShut {NoStop}%
\bibitem [{\citenamefont {Vuitton}\ \emph {et~al.}(2009)\citenamefont
  {Vuitton}, \citenamefont {Lavvas}, \citenamefont {Yelle}, \citenamefont
  {Galand}, \citenamefont {Wellbrock}, \citenamefont {Lewis}, \citenamefont
  {Coates},\ and\ \citenamefont {Wahlund}}]{VuittonNegativeionchemistry2009}%
  \BibitemOpen
  \bibfield  {author} {\bibinfo {author} {\bibfnamefont {V.}~\bibnamefont
  {Vuitton}}, \bibinfo {author} {\bibfnamefont {P.}~\bibnamefont {Lavvas}},
  \bibinfo {author} {\bibfnamefont {R.~V.}\ \bibnamefont {Yelle}}, \bibinfo
  {author} {\bibfnamefont {M.}~\bibnamefont {Galand}}, \bibinfo {author}
  {\bibfnamefont {A.}~\bibnamefont {Wellbrock}}, \bibinfo {author}
  {\bibfnamefont {G.~R.}\ \bibnamefont {Lewis}}, \bibinfo {author}
  {\bibfnamefont {A.~J.}\ \bibnamefont {Coates}},\ and\ \bibinfo {author}
  {\bibfnamefont {J.~E.}\ \bibnamefont {Wahlund}},\ }\bibfield  {title}
  {\bibinfo {title} {Negative ion chemistry in {{Titan}}'s upper atmosphere},\
  }\href {https://doi.org/10.1016/j.pss.2009.04.004} {\bibfield  {journal}
  {\bibinfo  {journal} {Planetary and Space Science}\ }\bibinfo {series}
  {Surfaces and {{Atmospheres}} of the {{Outer Planets}}, {{Their Satellites}}
  and {{Ring Systems}}: {{Part V}}},\ \textbf {\bibinfo {volume} {57}},\
  \bibinfo {pages} {1558} (\bibinfo {year} {2009})}\BibitemShut {NoStop}%
\bibitem [{\citenamefont {Sagawa}\ \emph {et~al.}(2005)\citenamefont {Sagawa},
  \citenamefont {Immel}, \citenamefont {Frey},\ and\ \citenamefont
  {Mende}}]{SagawaLongitudinalstructureequatorial2005}%
  \BibitemOpen
  \bibfield  {author} {\bibinfo {author} {\bibfnamefont {E.}~\bibnamefont
  {Sagawa}}, \bibinfo {author} {\bibfnamefont {T.~J.}\ \bibnamefont {Immel}},
  \bibinfo {author} {\bibfnamefont {H.~U.}\ \bibnamefont {Frey}},\ and\
  \bibinfo {author} {\bibfnamefont {S.~B.}\ \bibnamefont {Mende}},\ }\bibfield
  {title} {\bibinfo {title} {Longitudinal structure of the equatorial anomaly
  in the nighttime ionosphere observed by {{IMAGE}}/{{FUV}}},\ }\href
  {https://doi.org/10.1029/2004JA010848} {\bibfield  {journal} {\bibinfo
  {journal} {Journal of Geophysical Research (Space Physics)}\ }\textbf
  {\bibinfo {volume} {110}},\ \bibinfo {pages} {A11302} (\bibinfo {year}
  {2005})}\BibitemShut {NoStop}%
\bibitem [{\citenamefont {Barklem}\ \emph {et~al.}(2003)\citenamefont
  {Barklem}, \citenamefont {Belyaev},\ and\ \citenamefont
  {Asplund}}]{Barklem2003b}%
  \BibitemOpen
  \bibfield  {author} {\bibinfo {author} {\bibfnamefont {P.~S.}\ \bibnamefont
  {Barklem}}, \bibinfo {author} {\bibfnamefont {A.~K.}\ \bibnamefont
  {Belyaev}},\ and\ \bibinfo {author} {\bibfnamefont {M.}~\bibnamefont
  {Asplund}},\ }\bibfield  {title} {\bibinfo {title} {Inelastic {{H}}+{{Li}}
  and {{H}}{\textsuperscript{-}} +{{Li}}{\textsuperscript{+}} collisions and
  non-{{LTE Li I}} line formation in stellar atmospheres},\ }\href
  {https://doi.org/10.1051/0004-6361:20031199} {\bibfield  {journal} {\bibinfo
  {journal} {A\&A}\ }\textbf {\bibinfo {volume} {409}},\ \bibinfo {pages} {L1}
  (\bibinfo {year} {2003})}\BibitemShut {NoStop}%
\bibitem [{\citenamefont {Barklem}\ \emph {et~al.}(2021)\citenamefont
  {Barklem}, \citenamefont {Amarsi}, \citenamefont {Grumer}, \citenamefont
  {Eklund}, \citenamefont {Ros{\'e}n}, \citenamefont {Ji}, \citenamefont
  {Cederquist}, \citenamefont {Zettergren},\ and\ \citenamefont
  {Schmidt}}]{BarklemMutualNeutralizationLi2021}%
  \BibitemOpen
  \bibfield  {author} {\bibinfo {author} {\bibfnamefont {P.~S.}\ \bibnamefont
  {Barklem}}, \bibinfo {author} {\bibfnamefont {A.~M.}\ \bibnamefont {Amarsi}},
  \bibinfo {author} {\bibfnamefont {J.}~\bibnamefont {Grumer}}, \bibinfo
  {author} {\bibfnamefont {G.}~\bibnamefont {Eklund}}, \bibinfo {author}
  {\bibfnamefont {S.}~\bibnamefont {Ros{\'e}n}}, \bibinfo {author}
  {\bibfnamefont {M.}~\bibnamefont {Ji}}, \bibinfo {author} {\bibfnamefont
  {H.}~\bibnamefont {Cederquist}}, \bibinfo {author} {\bibfnamefont
  {H.}~\bibnamefont {Zettergren}},\ and\ \bibinfo {author} {\bibfnamefont
  {H.~T.}\ \bibnamefont {Schmidt}},\ }\bibfield  {title} {\bibinfo {title}
  {Mutual {{Neutralization}} in
  {{Li}}{\textsuperscript{+}}+{{H}}{\textsuperscript{-}}/{{D}}{\textsuperscript{-}}
  and
  {{Na}}{\textsuperscript{+}}+{{H}}{\textsuperscript{-}}/{{D}}{\textsuperscript{-}}
  {{Collisions}}: {{Implications}} of {{Experimental Results}} for
  {{Non}}-{{LTE Modeling}} of {{Stellar Spectra}}},\ }\href
  {https://doi.org/10.3847/1538-4357/abd5bd} {\bibfield  {journal} {\bibinfo
  {journal} {ApJ}\ }\textbf {\bibinfo {volume} {908}},\ \bibinfo {pages} {245}
  (\bibinfo {year} {2021})}\BibitemShut {NoStop}%
\bibitem [{\citenamefont {Fantz}\ and\ \citenamefont
  {W{\"u}nderlich}(2006)}]{Fantznoveldiagnostictechnique2006}%
  \BibitemOpen
  \bibfield  {author} {\bibinfo {author} {\bibfnamefont {U.}~\bibnamefont
  {Fantz}}\ and\ \bibinfo {author} {\bibfnamefont {D.}~\bibnamefont
  {W{\"u}nderlich}},\ }\bibfield  {title} {\bibinfo {title} {A novel diagnostic
  technique for {{H}}{\textsuperscript{-}}({{D}}{\textsuperscript{-}})
  densities in negative hydrogen ion sources},\ }\href
  {https://doi.org/10.1088/1367-2630/8/12/301} {\bibfield  {journal} {\bibinfo
  {journal} {New J. Phys.}\ }\textbf {\bibinfo {volume} {8}},\ \bibinfo {pages}
  {301} (\bibinfo {year} {2006})}\BibitemShut {NoStop}%
\bibitem [{\citenamefont {Zener}(1932)}]{zener_nonadiabatic_1932}%
  \BibitemOpen
  \bibfield  {author} {\bibinfo {author} {\bibfnamefont {C.}~\bibnamefont
  {Zener}},\ }\bibfield  {title} {\bibinfo {title} {Non-{{Adiabatic Crossing}}
  of {{Energy Levels}}},\ }\href@noop {} {\bibfield  {journal} {\bibinfo
  {journal} {Proc. Roy. Soc. London Ser. A., Math. Phys. Sci.}\ }\textbf
  {\bibinfo {volume} {137}},\ \bibinfo {pages} {696} (\bibinfo {year}
  {1932})}\BibitemShut {NoStop}%
\bibitem [{\citenamefont {Bates}(1962)}]{Bates14TheoreticalTreatment1962}%
  \BibitemOpen
  \bibfield  {author} {\bibinfo {author} {\bibfnamefont {D.~R.}\ \bibnamefont
  {Bates}},\ }\bibfield  {title} {\bibinfo {title} {14 - {{Theoretical
  Treatment}} of {{Collisions}} between {{Atomic Systems}}},\ }in\ \href
  {https://doi.org/10.1016/B978-0-12-081450-3.50018-7} {\emph {\bibinfo
  {booktitle} {Pure and {{Applied Physics}}}}},\ \bibinfo {series} {Atomic and
  {{Molecular Processes}}}, Vol.~\bibinfo {volume} {13},\ \bibinfo {editor}
  {edited by\ \bibinfo {editor} {\bibfnamefont {D.~R.}\ \bibnamefont {Bates}}}\
  (\bibinfo  {publisher} {{Elsevier}},\ \bibinfo {year} {1962})\ pp.\ \bibinfo
  {pages} {549--621}\BibitemShut {NoStop}%
\bibitem [{\citenamefont {Scott}\ \emph {et~al.}(1991)\citenamefont {Scott},
  \citenamefont {Dalgarno},\ and\ \citenamefont
  {Morgan}}]{ScottExchangeenergyH21991}%
  \BibitemOpen
  \bibfield  {author} {\bibinfo {author} {\bibfnamefont {T.~C.}\ \bibnamefont
  {Scott}}, \bibinfo {author} {\bibfnamefont {A.}~\bibnamefont {Dalgarno}},\
  and\ \bibinfo {author} {\bibfnamefont {J.~D.}\ \bibnamefont {Morgan},
  \bibfnamefont {III}},\ }\bibfield  {title} {\bibinfo {title} {Exchange energy
  of {{H2}}(+) calculated from polarization perturbation theory and the
  {{Holstein}}-{{Herring}} method},\ }\href
  {https://doi.org/10.1103/PhysRevLett.67.1419} {\bibfield  {journal} {\bibinfo
   {journal} {Physical Review Letters}\ }\textbf {\bibinfo {volume} {67}},\
  \bibinfo {pages} {1419} (\bibinfo {year} {1991})}\BibitemShut {NoStop}%
\bibitem [{\citenamefont {Chibisov}\ and\ \citenamefont
  {Janev}(1988)}]{Chibisov1988}%
  \BibitemOpen
  \bibfield  {author} {\bibinfo {author} {\bibfnamefont {M.}~\bibnamefont
  {Chibisov}}\ and\ \bibinfo {author} {\bibfnamefont {R.}~\bibnamefont
  {Janev}},\ }\bibfield  {title} {\bibinfo {title} {Asymptotic exchange
  interactions in ion-atom systems},\ }\href
  {https://doi.org/10.1016/S0370-1573(98)90002-3} {\bibfield  {journal}
  {\bibinfo  {journal} {Phys. Rep.}\ }\textbf {\bibinfo {volume} {166}},\
  \bibinfo {pages} {1} (\bibinfo {year} {1988})}\BibitemShut {NoStop}%
\bibitem [{\citenamefont {Davidovi{\'c}}\ and\ \citenamefont
  {Janev}(1969)}]{DavidovicResonantChargeExchange1969}%
  \BibitemOpen
  \bibfield  {author} {\bibinfo {author} {\bibfnamefont {D.~M.}\ \bibnamefont
  {Davidovi{\'c}}}\ and\ \bibinfo {author} {\bibfnamefont {R.~K.}\ \bibnamefont
  {Janev}},\ }\bibfield  {title} {\bibinfo {title} {Resonant {{Charge
  Exchange}} of the {{Negative Ions}} in {{Slow Collisions}} with {{Atoms}}},\
  }\href {https://doi.org/10.1103/PhysRev.186.89} {\bibfield  {journal}
  {\bibinfo  {journal} {Phys. Rev.}\ }\textbf {\bibinfo {volume} {186}},\
  \bibinfo {pages} {89} (\bibinfo {year} {1969})}\BibitemShut {NoStop}%
\bibitem [{\citenamefont {Janev}\ and\ \citenamefont
  {Salin}(1972)}]{Janev1972}%
  \BibitemOpen
  \bibfield  {author} {\bibinfo {author} {\bibfnamefont {R.~K.}\ \bibnamefont
  {Janev}}\ and\ \bibinfo {author} {\bibfnamefont {A.}~\bibnamefont {Salin}},\
  }\bibfield  {title} {\bibinfo {title} {Exchange interaction between ionic and
  covalent states of two atoms at large distances},\ }\href
  {https://doi.org/10.1088/0022-3700/5/2/012} {\bibfield  {journal} {\bibinfo
  {journal} {J. Phys. B: At. Mol. Phys.}\ }\textbf {\bibinfo {volume} {5}},\
  \bibinfo {pages} {177} (\bibinfo {year} {1972})}\BibitemShut {NoStop}%
\bibitem [{\citenamefont {Janev}(1976{\natexlab{a}})}]{Janev1976}%
  \BibitemOpen
  \bibfield  {author} {\bibinfo {author} {\bibfnamefont {R.~K.}\ \bibnamefont
  {Janev}},\ }\bibfield  {title} {\bibinfo {title} {On the long-range
  configuration interaction between ionic and covalent states},\ }\href
  {https://doi.org/10.1063/1.432473} {\bibfield  {journal} {\bibinfo  {journal}
  {J. Chem. Phys.}\ }\textbf {\bibinfo {volume} {64}},\ \bibinfo {pages} {1891}
  (\bibinfo {year} {1976}{\natexlab{a}})}\BibitemShut {NoStop}%
\bibitem [{\citenamefont {Firsov}(1951)}]{Firsov1951}%
  \BibitemOpen
  \bibfield  {author} {\bibinfo {author} {\bibfnamefont {O.~B.}\ \bibnamefont
  {Firsov}},\ }\href@noop {} {\bibfield  {journal} {\bibinfo  {journal} {Zh.
  Eksp. Teor. Fiz.}\ }\textbf {\bibinfo {volume} {21}},\ \bibinfo {pages}
  {1001} (\bibinfo {year} {1951})}\BibitemShut {NoStop}%
\bibitem [{\citenamefont
  {Holstein}(1952)}]{HolsteinMobilitiesPositiveIons1952}%
  \BibitemOpen
  \bibfield  {author} {\bibinfo {author} {\bibfnamefont {T.}~\bibnamefont
  {Holstein}},\ }\bibfield  {title} {\bibinfo {title} {Mobilities of {{Positive
  Ions}} in their {{Parent Gases}}},\ }\href
  {https://doi.org/10.1021/j150499a004} {\bibfield  {journal} {\bibinfo
  {journal} {J. Phys. Chem.}\ }\textbf {\bibinfo {volume} {56}},\ \bibinfo
  {pages} {832} (\bibinfo {year} {1952})}\BibitemShut {NoStop}%
\bibitem [{\citenamefont {Herring}(1962)}]{Herring1962}%
  \BibitemOpen
  \bibfield  {author} {\bibinfo {author} {\bibfnamefont {C.}~\bibnamefont
  {Herring}},\ }\bibfield  {title} {\bibinfo {title} {Critique of the
  {{Heitler}}-{{London Method}} of {{Calculating Spin Couplings}} at {{Large
  Distances}}},\ }\href {https://doi.org/10.1103/RevModPhys.34.631} {\bibfield
  {journal} {\bibinfo  {journal} {Rev. Mod. Phys.}\ }\textbf {\bibinfo {volume}
  {34}},\ \bibinfo {pages} {631} (\bibinfo {year} {1962})}\BibitemShut
  {NoStop}%
\bibitem [{\citenamefont {Landau}\ and\ \citenamefont
  {Lifshitz}(1965)}]{landau_quantum_1965}%
  \BibitemOpen
  \bibfield  {author} {\bibinfo {author} {\bibfnamefont {L.~D.}\ \bibnamefont
  {Landau}}\ and\ \bibinfo {author} {\bibfnamefont {E.~M.}\ \bibnamefont
  {Lifshitz}},\ }\href@noop {} {\emph {\bibinfo {title} {Quantum Mechanics}}}\
  (\bibinfo  {publisher} {{Pergamon Press}},\ \bibinfo {year}
  {1965})\BibitemShut {NoStop}%
\bibitem [{\citenamefont {Smirnov}(1965)}]{smirnov_formation_1965}%
  \BibitemOpen
  \bibfield  {author} {\bibinfo {author} {\bibfnamefont {B.~M.}\ \bibnamefont
  {Smirnov}},\ }\bibfield  {title} {\bibinfo {title} {Formation and {{Decay}}
  of {{Negative Ions}}},\ }\href@noop {} {\bibfield  {journal} {\bibinfo
  {journal} {Sov. Phys. Dok.}\ }\textbf {\bibinfo {volume} {10}},\ \bibinfo
  {pages} {218} (\bibinfo {year} {1965})}\BibitemShut {NoStop}%
\bibitem [{\citenamefont {Barklem}(2016)}]{barklem_excitation_2016}%
  \BibitemOpen
  \bibfield  {author} {\bibinfo {author} {\bibfnamefont {P.~S.}\ \bibnamefont
  {Barklem}},\ }\bibfield  {title} {\bibinfo {title} {Excitation and charge
  transfer in low-energy hydrogen-atom collisions with neutral atoms:
  {{Theory}}, comparisons, and application to {{Ca}}},\ }\href
  {https://doi.org/10.1103/PhysRevA.93.042705} {\bibfield  {journal} {\bibinfo
  {journal} {Phys. Rev. A}\ }\textbf {\bibinfo {volume} {93}},\ \bibinfo
  {pages} {042705} (\bibinfo {year} {2016})}\BibitemShut {NoStop}%
\bibitem [{\citenamefont {Barklem}(2017)}]{barklem_erratum:_2017}%
  \BibitemOpen
  \bibfield  {author} {\bibinfo {author} {\bibfnamefont {P.~S.}\ \bibnamefont
  {Barklem}},\ }\bibfield  {title} {\bibinfo {title} {Erratum: {{Excitation}}
  and charge transfer in low-energy hydrogen-atom collisions with neutral
  atoms: {{Theory}}, comparisons, and application to {{Ca}} [{{Phys}}. {{Rev}}.
  {{A}} 93, 042705 (2016)]},\ }\href
  {https://doi.org/10.1103/PhysRevA.95.069906} {\bibfield  {journal} {\bibinfo
  {journal} {Phys. Rev. A}\ }\textbf {\bibinfo {volume} {95}},\ \bibinfo
  {pages} {069906(E)} (\bibinfo {year} {2017})}\BibitemShut {NoStop}%
\bibitem [{\citenamefont {Janev}\ and\ \citenamefont
  {Radulovic}(1978)}]{Janev1978}%
  \BibitemOpen
  \bibfield  {author} {\bibinfo {author} {\bibfnamefont {R.~K.}\ \bibnamefont
  {Janev}}\ and\ \bibinfo {author} {\bibfnamefont {Z.~M.}\ \bibnamefont
  {Radulovic}},\ }\bibfield  {title} {\bibinfo {title} {Ion-ion recombination
  and ion-pair formation processes in alkali-hydrogen diatomic systems},\
  }\href {https://doi.org/10.1103/PhysRevA.17.889} {\bibfield  {journal}
  {\bibinfo  {journal} {Phys. Rev. A}\ }\textbf {\bibinfo {volume} {17}},\
  \bibinfo {pages} {889} (\bibinfo {year} {1978})}\BibitemShut {NoStop}%
\bibitem [{\citenamefont {Janev}\ \emph {et~al.}(2006)\citenamefont {Janev},
  \citenamefont {Liu}, \citenamefont {Wang},\ and\ \citenamefont
  {Yan}}]{Janev2006}%
  \BibitemOpen
  \bibfield  {author} {\bibinfo {author} {\bibfnamefont {R.~K.}\ \bibnamefont
  {Janev}}, \bibinfo {author} {\bibfnamefont {C.~L.}\ \bibnamefont {Liu}},
  \bibinfo {author} {\bibfnamefont {J.~G.}\ \bibnamefont {Wang}},\ and\
  \bibinfo {author} {\bibfnamefont {J.}~\bibnamefont {Yan}},\ }\bibfield
  {title} {\bibinfo {title} {Mutual neutralization of
  {{H}}{\textsubscript{3}}{\textsuperscript{+}} and {{H}}{\textsuperscript{-}}
  ions in slow collisions},\ }\href {https://doi.org/10.1209/epl/i2005-10576-1}
  {\bibfield  {journal} {\bibinfo  {journal} {EPL}\ }\textbf {\bibinfo {volume}
  {74}},\ \bibinfo {pages} {616} (\bibinfo {year} {2006})}\BibitemShut
  {NoStop}%
\bibitem [{\citenamefont {{Jian-Guo}}\ \emph {et~al.}(2006)\citenamefont
  {{Jian-Guo}}, \citenamefont {{Chun-Lei}}, \citenamefont {Janev},
  \citenamefont {Jun},\ and\ \citenamefont
  {{Jian-Rong}}}]{Jian-GuoMutualrecombinationslow2006}%
  \BibitemOpen
  \bibfield  {author} {\bibinfo {author} {\bibfnamefont {W.}~\bibnamefont
  {{Jian-Guo}}}, \bibinfo {author} {\bibfnamefont {L.}~\bibnamefont
  {{Chun-Lei}}}, \bibinfo {author} {\bibfnamefont {R.~K.}\ \bibnamefont
  {Janev}}, \bibinfo {author} {\bibfnamefont {Y.}~\bibnamefont {Jun}},\ and\
  \bibinfo {author} {\bibfnamefont {S.}~\bibnamefont {{Jian-Rong}}},\
  }\bibfield  {title} {\bibinfo {title} {Mutual recombination in slow
  {{Si}}{\textsuperscript{+}} + {{H}}{\textsuperscript{-}} collisions},\ }\href
  {https://doi.org/10.1088/1009-1963/15/11/032} {\bibfield  {journal} {\bibinfo
   {journal} {Chinese Phys.}\ }\textbf {\bibinfo {volume} {15}},\ \bibinfo
  {pages} {2651} (\bibinfo {year} {2006})}\BibitemShut {NoStop}%
\bibitem [{\citenamefont {Miyano~Hedberg}\ \emph {et~al.}(2014)\citenamefont
  {Miyano~Hedberg}, \citenamefont {Nkambule},\ and\ \citenamefont
  {Larson}}]{MiyanoHedberg2014}%
  \BibitemOpen
  \bibfield  {author} {\bibinfo {author} {\bibfnamefont {H.}~\bibnamefont
  {Miyano~Hedberg}}, \bibinfo {author} {\bibfnamefont {S.}~\bibnamefont
  {Nkambule}},\ and\ \bibinfo {author} {\bibfnamefont {{\AA}.}~\bibnamefont
  {Larson}},\ }\bibfield  {title} {\bibinfo {title} {Landau\textendash{{Zener}}
  studies of mutual neutralization in collisions of
  {{H}}{\textsuperscript{+}}/{{H}}{\textsuperscript{-}} and
  {{Be}}{\textsuperscript{+}}/{{H}}{\textsuperscript{-}}},\ }\href
  {https://doi.org/10.1088/0953-4075/47/22/225206} {\bibfield  {journal}
  {\bibinfo  {journal} {J. Phys. B: At. Mol. Opt. Phys.}\ }\textbf {\bibinfo
  {volume} {47}},\ \bibinfo {pages} {225206} (\bibinfo {year}
  {2014})}\BibitemShut {NoStop}%
\bibitem [{\citenamefont {Zhou}\ and\ \citenamefont
  {Dickinson}(1997)}]{zhou_mutual_1997}%
  \BibitemOpen
  \bibfield  {author} {\bibinfo {author} {\bibfnamefont {X.}~\bibnamefont
  {Zhou}}\ and\ \bibinfo {author} {\bibfnamefont {A.~S.}\ \bibnamefont
  {Dickinson}},\ }\bibfield  {title} {\bibinfo {title} {Mutual neutralisation
  of {{N}}{\textsuperscript{+}} by {{O}}{\textsuperscript{-}} and
  {{O}}{\textsuperscript{+}} by {{O}}{\textsuperscript{-}}},\ }\href
  {https://doi.org/10.1016/S0168-583X(96)00922-6} {\bibfield  {journal}
  {\bibinfo  {journal} {Nuc. Inst. Meth. in Phys. Res. B.}\ }\textbf {\bibinfo
  {volume} {124}},\ \bibinfo {pages} {5} (\bibinfo {year} {1997})}\BibitemShut
  {NoStop}%
\bibitem [{\citenamefont {{de Ruette}}\ \emph {et~al.}(2018)\citenamefont {{de
  Ruette}}, \citenamefont {Dochain}, \citenamefont {Launoy}, \citenamefont
  {Nascimento}, \citenamefont {Kaminska}, \citenamefont {Stockett},
  \citenamefont {Vaeck}, \citenamefont {Schmidt}, \citenamefont {Cederquist},\
  and\ \citenamefont {Urbain}}]{deRuetteMutualNeutralizationSubthermal2018}%
  \BibitemOpen
  \bibfield  {author} {\bibinfo {author} {\bibfnamefont {N.}~\bibnamefont {{de
  Ruette}}}, \bibinfo {author} {\bibfnamefont {A.}~\bibnamefont {Dochain}},
  \bibinfo {author} {\bibfnamefont {T.}~\bibnamefont {Launoy}}, \bibinfo
  {author} {\bibfnamefont {R.~F.}\ \bibnamefont {Nascimento}}, \bibinfo
  {author} {\bibfnamefont {M.}~\bibnamefont {Kaminska}}, \bibinfo {author}
  {\bibfnamefont {M.~H.}\ \bibnamefont {Stockett}}, \bibinfo {author}
  {\bibfnamefont {N.}~\bibnamefont {Vaeck}}, \bibinfo {author} {\bibfnamefont
  {H.~T.}\ \bibnamefont {Schmidt}}, \bibinfo {author} {\bibfnamefont
  {H.}~\bibnamefont {Cederquist}},\ and\ \bibinfo {author} {\bibfnamefont
  {X.}~\bibnamefont {Urbain}},\ }\bibfield  {title} {\bibinfo {title} {Mutual
  {{Neutralization}} of {{O}}{\textsuperscript{-}} with
  {{O}}{\textsuperscript{+}} and {{N}}{\textsuperscript{+}} at {{Subthermal
  Collision Energies}}},\ }\href
  {https://doi.org/10.1103/PhysRevLett.121.083401} {\bibfield  {journal}
  {\bibinfo  {journal} {Physical Review Letters}\ }\textbf {\bibinfo {volume}
  {121}},\ \bibinfo {pages} {083401} (\bibinfo {year} {2018})}\BibitemShut
  {NoStop}%
\bibitem [{\citenamefont {Landau}(1932{\natexlab{a}})}]{landau_1932}%
  \BibitemOpen
  \bibfield  {author} {\bibinfo {author} {\bibfnamefont {L.~D.}\ \bibnamefont
  {Landau}},\ }\href@noop {} {\bibfield  {journal} {\bibinfo  {journal}
  {Sowietunion}\ }\textbf {\bibinfo {volume} {1}},\ \bibinfo {pages} {88}
  (\bibinfo {year} {1932}{\natexlab{a}})}\BibitemShut {NoStop}%
\bibitem [{\citenamefont {Landau}(1932{\natexlab{b}})}]{landau_1932-1}%
  \BibitemOpen
  \bibfield  {author} {\bibinfo {author} {\bibfnamefont {L.~D.}\ \bibnamefont
  {Landau}},\ }\href@noop {} {\bibfield  {journal} {\bibinfo  {journal}
  {Sowietunion}\ }\textbf {\bibinfo {volume} {2}},\ \bibinfo {pages} {46}
  (\bibinfo {year} {1932}{\natexlab{b}})}\BibitemShut {NoStop}%
\bibitem [{\citenamefont {Andreev}(1973)}]{AndreevExchangeinteractiontwo1973}%
  \BibitemOpen
  \bibfield  {author} {\bibinfo {author} {\bibfnamefont {E.~A.}\ \bibnamefont
  {Andreev}},\ }\bibfield  {title} {\bibinfo {title} {Exchange interaction
  between two different atoms at large distances},\ }\href
  {https://doi.org/10.1007/BF01209766} {\bibfield  {journal} {\bibinfo
  {journal} {Theoret. Chim. Acta}\ }\textbf {\bibinfo {volume} {30}},\ \bibinfo
  {pages} {191} (\bibinfo {year} {1973})}\BibitemShut {NoStop}%
\bibitem [{\citenamefont {Grice}\ and\ \citenamefont
  {Herschbach}(1974)}]{Grice1974}%
  \BibitemOpen
  \bibfield  {author} {\bibinfo {author} {\bibfnamefont {R.}~\bibnamefont
  {Grice}}\ and\ \bibinfo {author} {\bibfnamefont {D.~R.}\ \bibnamefont
  {Herschbach}},\ }\bibfield  {title} {\bibinfo {title} {Long-range
  configuration interaction of ionic and covalent states},\ }\href
  {https://doi.org/10.1080/00268977400100131} {\bibfield  {journal} {\bibinfo
  {journal} {Mol. Phys.}\ }\textbf {\bibinfo {volume} {27}},\ \bibinfo {pages}
  {159} (\bibinfo {year} {1974})}\BibitemShut {NoStop}%
\bibitem [{\citenamefont {Nikitin}\ and\ \citenamefont
  {Umanskii}(1984)}]{Nikitin1984}%
  \BibitemOpen
  \bibfield  {author} {\bibinfo {author} {\bibfnamefont {E.~E.}\ \bibnamefont
  {Nikitin}}\ and\ \bibinfo {author} {\bibfnamefont {S.~I.}\ \bibnamefont
  {Umanskii}},\ }\href@noop {} {\emph {\bibinfo {title} {Theory of Slow Atomic
  Collisions}}},\ Vol.~\bibinfo {volume} {30}\ (\bibinfo  {publisher}
  {{Springer}},\ \bibinfo {address} {{Berlin and New York}},\ \bibinfo {year}
  {1984})\BibitemShut {NoStop}%
\bibitem [{\citenamefont {Olver}\ and\ \citenamefont {{National Institute of
  Standards {and} Technology
  (U.S.)}}(2010)}]{OlverNISThandbookmathematical2010}%
  \BibitemOpen
  \bibinfo {editor} {\bibfnamefont {F.~W.~J.}\ \bibnamefont {Olver}}\ and\
  \bibinfo {editor} {\bibnamefont {{National Institute of Standards {and}
  Technology (U.S.)}}},\ eds.,\ \href@noop {} {\emph {\bibinfo {title}
  {{{NIST}} Handbook of Mathematical Functions}}}\ (\bibinfo  {publisher}
  {{Cambridge University Press : NIST}},\ \bibinfo {address} {{Cambridge ; New
  York}},\ \bibinfo {year} {2010})\BibitemShut {NoStop}%
\bibitem [{\citenamefont {Hartree}(1928)}]{HartreeWaveMechanicsAtom1928}%
  \BibitemOpen
  \bibfield  {author} {\bibinfo {author} {\bibfnamefont {D.~R.}\ \bibnamefont
  {Hartree}},\ }\bibfield  {title} {\bibinfo {title} {The {{Wave Mechanics}} of
  an {{Atom}} with a {{Non}}-{{Coulomb Central Field}}. {{Part I}}. {{Theory}}
  and {{Methods}}},\ }\href {https://doi.org/10.1017/S0305004100011919}
  {\bibfield  {journal} {\bibinfo  {journal} {Mathematical Proceedings of the
  Cambridge Philosophical Society}\ }\textbf {\bibinfo {volume} {24}},\
  \bibinfo {pages} {89} (\bibinfo {year} {1928})}\BibitemShut {NoStop}%
\bibitem [{\citenamefont {Bates}\ and\ \citenamefont
  {Damgaard}(1949)}]{Bates1949}%
  \BibitemOpen
  \bibfield  {author} {\bibinfo {author} {\bibfnamefont {D.~R.}\ \bibnamefont
  {Bates}}\ and\ \bibinfo {author} {\bibfnamefont {A.}~\bibnamefont
  {Damgaard}},\ }\bibfield  {title} {\bibinfo {title} {The {{Calculation}} of
  the {{Absolute Strengths}} of {{Spectral Lines}}},\ }\href
  {https://doi.org/10.1098/rsta.1949.0006} {\bibfield  {journal} {\bibinfo
  {journal} {Philosophical Transactions of the Royal Society A: Mathematical,
  Physical and Engineering Sciences}\ }\textbf {\bibinfo {volume} {242}},\
  \bibinfo {pages} {101} (\bibinfo {year} {1949})}\BibitemShut {NoStop}%
\bibitem [{\citenamefont {Seaton}(1958)}]{Seaton1958a}%
  \BibitemOpen
  \bibfield  {author} {\bibinfo {author} {\bibfnamefont {M.~J.}\ \bibnamefont
  {Seaton}},\ }\bibfield  {title} {\bibinfo {title} {The {{Quantum Defect
  Method}}},\ }\href@noop {} {\bibfield  {journal} {\bibinfo  {journal}
  {Monthly Notices of the Royal Astronomical Society}\ }\textbf {\bibinfo
  {volume} {118}},\ \bibinfo {pages} {504} (\bibinfo {year}
  {1958})}\BibitemShut {NoStop}%
\bibitem [{\citenamefont {Smirnov}(2001)}]{SmirnovAtomicstructureresonant2001}%
  \BibitemOpen
  \bibfield  {author} {\bibinfo {author} {\bibfnamefont {B.~M.}\ \bibnamefont
  {Smirnov}},\ }\bibfield  {title} {\bibinfo {title} {Atomic structure and the
  resonant charge exchange process},\ }\href
  {https://doi.org/10.1070/PU2001v044n03ABEH000826} {\bibfield  {journal}
  {\bibinfo  {journal} {Phys.-Usp.}\ }\textbf {\bibinfo {volume} {44}},\
  \bibinfo {pages} {221} (\bibinfo {year} {2001})}\BibitemShut {NoStop}%
\bibitem [{\citenamefont
  {Janev}(1976{\natexlab{b}})}]{JanevNonadiabaticTransitionsIonic1976}%
  \BibitemOpen
  \bibfield  {author} {\bibinfo {author} {\bibfnamefont {R.~K.}\ \bibnamefont
  {Janev}},\ }\bibfield  {title} {\bibinfo {title} {Nonadiabatic {{Transitions
  Between Ionic}} and {{Covalent States}}},\ }in\ \href
  {https://doi.org/10.1016/S0065-2199(08)60041-X} {\emph {\bibinfo {booktitle}
  {Advances in {{Atomic}} and {{Molecular Physics}}}}},\ Vol.~\bibinfo {volume}
  {12},\ \bibinfo {editor} {edited by\ \bibinfo {editor} {\bibfnamefont
  {D.~R.}\ \bibnamefont {Bates}}\ and\ \bibinfo {editor} {\bibfnamefont
  {B.}~\bibnamefont {Bederson}}}\ (\bibinfo  {publisher} {{Academic Press}},\
  \bibinfo {year} {1976})\ pp.\ \bibinfo {pages} {1--37}\BibitemShut {NoStop}%
\bibitem [{\citenamefont {Adelman}\ and\ \citenamefont
  {Herschbach}(1977)}]{Adelman1977}%
  \BibitemOpen
  \bibfield  {author} {\bibinfo {author} {\bibfnamefont {S.~A.}\ \bibnamefont
  {Adelman}}\ and\ \bibinfo {author} {\bibfnamefont {D.~R.}\ \bibnamefont
  {Herschbach}},\ }\bibfield  {title} {\bibinfo {title} {Asymptotic
  approximation for ionic-covalent configuration mixing in hydrogen and alkali
  hydrides},\ }\href {https://doi.org/10.1080/00268977700100731} {\bibfield
  {journal} {\bibinfo  {journal} {Mol. Phys.}\ }\textbf {\bibinfo {volume}
  {33}},\ \bibinfo {pages} {793} (\bibinfo {year} {1977})}\BibitemShut
  {NoStop}%
\bibitem [{\citenamefont {Anstee}(1992)}]{Anstee1992}%
  \BibitemOpen
  \bibfield  {author} {\bibinfo {author} {\bibfnamefont {S.~D.}\ \bibnamefont
  {Anstee}},\ }\emph {\bibinfo {title} {The {{Collisional Broadening}} of
  {{Alkali Spectral Lines}} by {{Atomic Hydrogen}}}},\ \href@noop {} {Ph.D.
  thesis},\ \bibinfo  {school} {The University of Queensland} (\bibinfo {year}
  {1992})\BibitemShut {NoStop}%
\bibitem [{\citenamefont {Smirnov}(2003)}]{SmirnovPhysicsatomsions2003}%
  \BibitemOpen
  \bibfield  {author} {\bibinfo {author} {\bibfnamefont {B.~M.}\ \bibnamefont
  {Smirnov}},\ }\href@noop {} {\emph {\bibinfo {title} {Physics of Atoms and
  Ions}}},\ Graduate Texts in Contemporary Physics\ (\bibinfo  {publisher}
  {{Springer}},\ \bibinfo {address} {{New York}},\ \bibinfo {year}
  {2003})\BibitemShut {NoStop}%
\bibitem [{\citenamefont {Patil}(2013)}]{PatilAsymptoticmethodsquantum2013}%
  \BibitemOpen
  \bibfield  {author} {\bibinfo {author} {\bibfnamefont {S.~H.}\ \bibnamefont
  {Patil}},\ }\href@noop {} {\emph {\bibinfo {title} {Asymptotic Methods in
  Quantum Mechanics: Application to Atoms, Molecules and Nuclei.}}}\ (\bibinfo
  {publisher} {{Springer}},\ \bibinfo {address} {{Place of publication not
  identified}},\ \bibinfo {year} {2013})\BibitemShut {NoStop}%
\bibitem [{\citenamefont
  {Patil}(1995)}]{PatilAsymptoticbehaviourwavefunctions1995}%
  \BibitemOpen
  \bibfield  {author} {\bibinfo {author} {\bibfnamefont {S.}~\bibnamefont
  {Patil}},\ }\bibfield  {title} {\bibinfo {title} {Asymptotic behaviour of
  wavefunctions: Its applications to properties of atoms, molecules and
  nuclei},\ }\href {https://doi.org/10.1088/0143-0807/16/1/005} {\bibfield
  {journal} {\bibinfo  {journal} {Eur. J. Phys.}\ }\textbf {\bibinfo {volume}
  {16}},\ \bibinfo {pages} {25} (\bibinfo {year} {1995})}\BibitemShut {NoStop}%
\bibitem [{\citenamefont {Lykke}\ \emph {et~al.}(1991)\citenamefont {Lykke},
  \citenamefont {Murray},\ and\ \citenamefont
  {Lineberger}}]{LykkeThresholdphotodetachment1991}%
  \BibitemOpen
  \bibfield  {author} {\bibinfo {author} {\bibfnamefont {K.~R.}\ \bibnamefont
  {Lykke}}, \bibinfo {author} {\bibfnamefont {K.~K.}\ \bibnamefont {Murray}},\
  and\ \bibinfo {author} {\bibfnamefont {W.~C.}\ \bibnamefont {Lineberger}},\
  }\bibfield  {title} {\bibinfo {title} {Threshold photodetachment of
  {{H}}{\textsuperscript{-}}},\ }\href
  {https://doi.org/10.1103/PhysRevA.43.6104} {\bibfield  {journal} {\bibinfo
  {journal} {Phys. Rev. A}\ }\textbf {\bibinfo {volume} {43}},\ \bibinfo
  {pages} {6104} (\bibinfo {year} {1991})}\BibitemShut {NoStop}%
\bibitem [{\citenamefont {Smirnov}(1973)}]{smirnov_asymptotic_1973}%
  \BibitemOpen
  \bibfield  {author} {\bibinfo {author} {\bibfnamefont {B.~M.}\ \bibnamefont
  {Smirnov}},\ }\href@noop {} {\emph {\bibinfo {title} {Asymptotic {{Methods}}
  in the {{Theory}} of {{Atomic Collisions}} (in {{Russian}})}}}\ (\bibinfo
  {publisher} {{Nauka}},\ \bibinfo {address} {{Moscow}},\ \bibinfo {year}
  {1973})\BibitemShut {NoStop}%
\bibitem [{\citenamefont {Chandrasekhar}(1944)}]{chandrasekhar_remarks_1944}%
  \BibitemOpen
  \bibfield  {author} {\bibinfo {author} {\bibfnamefont {S.}~\bibnamefont
  {Chandrasekhar}},\ }\bibfield  {title} {\bibinfo {title} {Some {{Remarks}} on
  the {{Negative Hydrogen Ion}} and its {{Absorption Coefficient}}.},\ }\href
  {https://doi.org/10.1086/144654} {\bibfield  {journal} {\bibinfo  {journal}
  {The Astrophysical Journal}\ }\textbf {\bibinfo {volume} {100}},\ \bibinfo
  {pages} {176} (\bibinfo {year} {1944})}\BibitemShut {NoStop}%
\bibitem [{\citenamefont {Pekeris}(1958)}]{pekeris_ground_1958}%
  \BibitemOpen
  \bibfield  {author} {\bibinfo {author} {\bibfnamefont {C.~L.}\ \bibnamefont
  {Pekeris}},\ }\bibfield  {title} {\bibinfo {title} {Ground {{State}} of
  {{Two}}-{{Electron Atoms}}},\ }\href
  {https://doi.org/10.1103/PhysRev.112.1649} {\bibfield  {journal} {\bibinfo
  {journal} {Phys. Rev.}\ }\textbf {\bibinfo {volume} {112}},\ \bibinfo {pages}
  {1649} (\bibinfo {year} {1958})}\BibitemShut {NoStop}%
\bibitem [{\citenamefont {Ohmura}\ and\ \citenamefont
  {Ohmura}(1960)}]{Ohmura1960}%
  \BibitemOpen
  \bibfield  {author} {\bibinfo {author} {\bibfnamefont {T.}~\bibnamefont
  {Ohmura}}\ and\ \bibinfo {author} {\bibfnamefont {H.}~\bibnamefont
  {Ohmura}},\ }\bibfield  {title} {\bibinfo {title} {Electron-{{Hydrogen
  Scattering}} at {{Low Energies}}},\ }\href
  {https://doi.org/10.1103/PhysRev.118.154} {\bibfield  {journal} {\bibinfo
  {journal} {Physical Review}\ }\textbf {\bibinfo {volume} {118}},\ \bibinfo
  {pages} {154} (\bibinfo {year} {1960})}\BibitemShut {NoStop}%
\bibitem [{\citenamefont {Ohmura}\ \emph {et~al.}(1958)\citenamefont {Ohmura},
  \citenamefont {Hara},\ and\ \citenamefont
  {Yamanouchi}}]{OhmuraLowEnergyElectronHydrogen1958}%
  \BibitemOpen
  \bibfield  {author} {\bibinfo {author} {\bibfnamefont {T.}~\bibnamefont
  {Ohmura}}, \bibinfo {author} {\bibfnamefont {Y.}~\bibnamefont {Hara}},\ and\
  \bibinfo {author} {\bibfnamefont {T.}~\bibnamefont {Yamanouchi}},\ }\bibfield
   {title} {\bibinfo {title} {Low {{Energy Electron}}-{{Hydrogen
  Scattering}}},\ }\href {https://doi.org/10.1143/PTP.20.82} {\bibfield
  {journal} {\bibinfo  {journal} {Progress of Theoretical Physics}\ }\textbf
  {\bibinfo {volume} {20}},\ \bibinfo {pages} {82} (\bibinfo {year}
  {1958})}\BibitemShut {NoStop}%
\bibitem [{\citenamefont {Hart}\ and\ \citenamefont
  {Herzberg}(1957)}]{HartTwentyParameterEigenfunctionsEnergy1957}%
  \BibitemOpen
  \bibfield  {author} {\bibinfo {author} {\bibfnamefont {J.~F.}\ \bibnamefont
  {Hart}}\ and\ \bibinfo {author} {\bibfnamefont {G.}~\bibnamefont
  {Herzberg}},\ }\bibfield  {title} {\bibinfo {title} {Twenty-{{Parameter
  Eigenfunctions}} and {{Energy Values}} of the {{Ground States}} of {{He}} and
  {{He}}-{{Like Ions}}},\ }\href {https://doi.org/10.1103/PhysRev.106.79}
  {\bibfield  {journal} {\bibinfo  {journal} {Physical Review}\ }\textbf
  {\bibinfo {volume} {106}},\ \bibinfo {pages} {79} (\bibinfo {year}
  {1957})}\BibitemShut {NoStop}%
\bibitem [{\citenamefont
  {Hylleraas}(1929)}]{HylleraasNeueBerechnungEnergie1929}%
  \BibitemOpen
  \bibfield  {author} {\bibinfo {author} {\bibfnamefont {E.~A.}\ \bibnamefont
  {Hylleraas}},\ }\bibfield  {title} {\bibinfo {title} {{Neue Berechnung der
  Energie des Heliums im Grundzustande, sowie des tiefsten Terms von
  Ortho-Helium}},\ }\href {https://doi.org/10.1007/BF01375457} {\bibfield
  {journal} {\bibinfo  {journal} {Z. Physik}\ }\textbf {\bibinfo {volume}
  {54}},\ \bibinfo {pages} {347} (\bibinfo {year} {1929})}\BibitemShut
  {NoStop}%
\bibitem [{\citenamefont {Pan}\ \emph {et~al.}(2003)\citenamefont {Pan},
  \citenamefont {Sahni},\ and\ \citenamefont
  {Massa}}]{PanHylleraasCoordinates2003}%
  \BibitemOpen
  \bibfield  {author} {\bibinfo {author} {\bibfnamefont {X.-Y.}\ \bibnamefont
  {Pan}}, \bibinfo {author} {\bibfnamefont {V.}~\bibnamefont {Sahni}},\ and\
  \bibinfo {author} {\bibfnamefont {L.}~\bibnamefont {Massa}},\ }\bibfield
  {title} {\bibinfo {title} {On the {{Hylleraas Coordinates}}},\ }\href@noop {}
  {\bibfield  {journal} {\bibinfo  {journal} {arXiv:physics/0310128}\ }
  (\bibinfo {year} {2003})},\ \bibinfo {note} {comment: 9 pages, 4 figures, 1
  table},\ \Eprint {https://arxiv.org/abs/physics/0310128}
  {arXiv:physics/0310128} \BibitemShut {NoStop}%
\bibitem [{\citenamefont {Croft}\ \emph
  {et~al.}(1999{\natexlab{a}})\citenamefont {Croft}, \citenamefont
  {Dickinson},\ and\ \citenamefont {Gad{\'e}a}}]{Croft1999a}%
  \BibitemOpen
  \bibfield  {author} {\bibinfo {author} {\bibfnamefont {H.}~\bibnamefont
  {Croft}}, \bibinfo {author} {\bibfnamefont {A.~S.}\ \bibnamefont
  {Dickinson}},\ and\ \bibinfo {author} {\bibfnamefont {F.~X.}\ \bibnamefont
  {Gad{\'e}a}},\ }\bibfield  {title} {\bibinfo {title} {A theoretical study of
  mutual neutralization in
  {{Li}}{\textsuperscript{+}}+{{H}}{\textsuperscript{-}} collisions},\ }\href
  {https://doi.org/10.1088/0953-4075/32/1/008} {\bibfield  {journal} {\bibinfo
  {journal} {J. Phys. B: At. Mol. Opt. Phys.}\ }\textbf {\bibinfo {volume}
  {32}},\ \bibinfo {pages} {81} (\bibinfo {year}
  {1999}{\natexlab{a}})}\BibitemShut {NoStop}%
\bibitem [{\citenamefont {Croft}\ \emph
  {et~al.}(1999{\natexlab{b}})\citenamefont {Croft}, \citenamefont
  {Dickinson},\ and\ \citenamefont {Gad{\'e}a}}]{Croft1999}%
  \BibitemOpen
  \bibfield  {author} {\bibinfo {author} {\bibfnamefont {H.}~\bibnamefont
  {Croft}}, \bibinfo {author} {\bibfnamefont {A.~S.}\ \bibnamefont
  {Dickinson}},\ and\ \bibinfo {author} {\bibfnamefont {F.~X.}\ \bibnamefont
  {Gad{\'e}a}},\ }\bibfield  {title} {\bibinfo {title} {Rate coefficients for
  the {{Li}}{\textsuperscript{+}}/{{H}}{\textsuperscript{-}} and
  {{Li}}{\textsuperscript{-}}/{{H}}{\textsuperscript{+}} mutual neutralization
  reactions},\ }\href {https://doi.org/10.1046/j.1365-8711.1999.02346.x}
  {\bibfield  {journal} {\bibinfo  {journal} {MNRAS}\ }\textbf {\bibinfo
  {volume} {304}},\ \bibinfo {pages} {327} (\bibinfo {year}
  {1999}{\natexlab{b}})}\BibitemShut {NoStop}%
\bibitem [{\citenamefont {Dickinson}\ \emph {et~al.}(1999)\citenamefont
  {Dickinson}, \citenamefont {Poteau},\ and\ \citenamefont
  {Gad{\'e}a}}]{dickinson_initio_1999}%
  \BibitemOpen
  \bibfield  {author} {\bibinfo {author} {\bibfnamefont {A.~S.}\ \bibnamefont
  {Dickinson}}, \bibinfo {author} {\bibfnamefont {R.}~\bibnamefont {Poteau}},\
  and\ \bibinfo {author} {\bibfnamefont {F.~X.}\ \bibnamefont {Gad{\'e}a}},\
  }\bibfield  {title} {\bibinfo {title} {An ab initio study of mutual
  neutralization in {{Na}}{\textsuperscript{+}}+{{H}}{\textsuperscript{-}}
  collisions},\ }\href {https://doi.org/10.1088/0953-4075/32/23/303} {\bibfield
   {journal} {\bibinfo  {journal} {J. Phys. B: At. Mol. Opt. Phys.}\ }\textbf
  {\bibinfo {volume} {32}},\ \bibinfo {pages} {5451} (\bibinfo {year}
  {1999})}\BibitemShut {NoStop}%
\bibitem [{Note1()}]{Note1}%
  \BibitemOpen
  \bibinfo {note} {Exchange interaction is used in two senses here, in H$_2$
  the usual interaction due to antisymmetrization with respect to exchange of
  electron labels, and in H$_2^+$, due to the degeneracy with respect to
  exchange of position of the electron from one nucleus to the
  other.}\BibitemShut {Stop}%
\end{thebibliography}%

\appendix
\section{One-electron LCAO integrals for $l_c=l_i=0$.}
\label{sect:app}

\begin{widetext}
The matrix element $T_{ic}(R)$ can be derived analytically using the asymptotic atomic wavefunctions, eqns.~\ref{eq:wf_cov} and~\ref{eq:wf_ion} for the radial parts $\mathcal{R}_c$ and $\mathcal{R}_i$ of the asymptotic spatial atomic wavefunctions $\varphi_c$ and $\varphi_i$, respectively.  The simplest possible case is taken, where both $\varphi_c$ and $\varphi_i$ are spherically symmetric and $l_c=l_i=0$.  It is instructive to first consider the overlap, which can be shown to be
\begin{equation}
S_{ic}(R) = \langle \varphi_c |  \varphi_i \rangle = \frac{N_i}{\gamma_i R} \int_0^R dr \, r \, \mathcal{R}_{c}(r) e^{-\gamma_i R} \sinh(\gamma_i r) + \int_R^\infty dr \, r \, \mathcal{R}_{c}(r) e^{-\gamma_i r} \sinh(\gamma_i R) 
\label{eq:overlap}
\end{equation}
where $r=r_A$; see Fig.~\ref{fig:system}.  Performing these integrals with the aid of Mathematica we obtain in its most compact form
\begin{multline}
S_{ic}(R) = -\frac{N_c N_i}{2 \gamma_i R} ((\gamma_c-\gamma_i) (\gamma_c+\gamma_i))^{-\frac{\gamma_c+1}{\gamma_c}}
   e^{-\gamma_i R} \left\{(\gamma_c-\gamma_i)^{\frac{1}{\gamma_c}+1} \left(e^{2 \gamma_i R} \Gamma
   \left(1+\frac{1}{\gamma_c},(\gamma_c+\gamma_i) R\right)-\Gamma 
   \left(1+\frac{1}{\gamma_c}\right)\right) \right. \\
   \left. +(\gamma_c+\gamma_i)^{\frac{1}{\gamma_c}+1} \left(\Gamma
   \left(1+\frac{1}{\gamma_c}\right)-\Gamma \left(1+\frac{1}{\gamma_c},(\gamma_c-\gamma_i)
   R\right)\right)\right\},
\end{multline}
where $\Gamma(z)$ is the Euler gamma function, and $\Gamma(a,z)$ is the incomplete gamma function; see e.g. Ref~\cite{OlverNISThandbookmathematical2010}.
Similarly, $T_{ic}(R)$ can be evaluated simply by adding the operator $T= 2 (\frac{1}{r_A}   - \frac{1}{R})$ into eqn.~\ref{eq:overlap}.  Evaluating these integrals gives, in its most compact form
\begin{multline}
T_{ic}(R) =  -\frac{N_c N_i}{\gamma_c \gamma_i R^2} ((\gamma_c-\gamma_i) (\gamma_c+\gamma_i))^{-\frac{\gamma_c+1}{\gamma_c}}
   e^{-\gamma_i R} \Bigg\{ 2 \gamma_c \gamma_i e^{R (\gamma_i-\gamma_c)} (R (\gamma_c-\gamma_i)
   (\gamma_c+\gamma_i))^{\frac{1}{\gamma_c}}  \\
   \left. -\left((\gamma_c-\gamma_i)^{\frac{1}{\gamma_c}+1} (\gamma_c R
   (\gamma_c+\gamma_i)-1) \left(\Gamma \left(\frac{1}{\gamma_c}\right)-e^{2 \gamma_i R} \Gamma
   \left(\frac{1}{\gamma_c},(\gamma_c+\gamma_i)
   R\right)\right)\right) \right. \\ 
    +(\gamma_c+\gamma_i)^{\frac{1}{\gamma_c}+1} (\gamma_c R (\gamma_c-\gamma_i)-1)
   \left(\Gamma \left(\frac{1}{\gamma_c}\right)-\Gamma \left(\frac{1}{\gamma_c},(\gamma_c-\gamma_i)
   R\right)\right) \Bigg\}.
\end{multline}   
\end{widetext}

\end{document}